\documentclass{article}
\usepackage{spconf,amsmath,graphicx}
\usepackage{amsfonts}
\usepackage{multirow}
\usepackage{booktabs}
\usepackage[table]{xcolor}
\usepackage{caption}
\usepackage{url}

\title{END-TO-END NEURAL SYSTEMS FOR AUTOMATIC CHILDREN SPEECH RECOGNITION: AN EMPIRICAL STUDY}
\name{Prashanth Gurunath Shivakumar\thanks{\texttt{email: pgurunat@usc.edu; shri@sipi.usc.edu}}, Shrikanth Narayanan}
\address{University of Southern California, Los Angeles, California 90089, USA}

\begin{document}
%
\maketitle
\begin{abstract}
A key desiderata for inclusive and accessible speech recognition technology is ensuring its robust performance to children's speech. 
Notably, this includes the rapidly advancing neural network based end-to-end speech recognition systems. 
Children speech recognition is more challenging due to the larger intra-inter speaker variability in terms of acoustic and linguistic characteristics compared to adult speech. Furthermore, the lack of adequate and appropriate children speech resources adds to the challenge of designing robust end-to-end neural architectures.
This study provides a critical assessment of automatic children speech recognition through an empirical study of contemporary state-of-the-art end-to-end speech recognition systems.
Insights are provided on the aspects of training data requirements, adaptation on children data, and the effect of children age, utterance lengths, different architectures and loss functions for end-to-end systems and role of language models on the speech recognition performance.
\end{abstract}
\begin{keywords}
Children Speech Recognition, End-to-End Speech Recognition, Residual Network, Time Depth Separable Convolutional Network, Transformer
\end{keywords}
\section{Introduction}
\label{sec:intro}

Creating speech and spoken language technologies (SLT) that are inclusive and broadly accessible need to ensure that they offer robust performance to children speech. Beyond supporting an important potential segment of users of applications involving conversational interfaces \cite{Narayanan2002Creatingconversationalinterfacesfor} such as for entertainment, interactive gaming, education and learning, such technologies can enable novel child-centric possibilities in support of diagnosis and treatment for a variety of developmental disorders and health conditions \cite{Bone2017SignalProcessingandMachine,bone2017behavioral}.  
However, the inclusion of the children population in SLT research and development has been lagging behind within the exciting realm of rapid development and deployment of these technologies mainly for adult population, underscoring an unmet need.

Automatic speech recognition (ASR) is a core SLT technology, and has witnessed accelerated advances since the advent of deep learning.
Early attempts at incorporating deep learning into ASR replaced Gaussian mixture models (GMM) with DNN \cite{dahl2011context}.
The objective of the DNN is to produce a distribution over senones given the input acoustic feature frames.
Such a system requires the alignments obtained from the GMM-based hidden markov models (HMM) for training purposes.
\cite{graves2006connectionist} introduced connectionist temporal classification (CTC) for sequence data labeling with recurrent neural networks which eliminates the need of pre-computed alignments and subsequent processing by computing the probability distribution over all the possible label sequences given the input signal sequence.
Alternatively to the CTC, sequence to sequence learning was introduced to compute the mapping between variable length sequences \cite{sutskever2014sequence}; and attention-based sequence to sequence models proved feasible for end-to-end trainable speech recognition systems \cite{chorowski2015attention}.
The attention mechanism is able to implicitly calculate the alignments between the sequence of input speech feature frames and the output text sequence provided with large amounts of training data. 

Several different end-to-end DNN architectures for ASR have been proposed with CTC and sequence-to-sequence learning frameworks.
RNN based architectures \cite{chorowski2015attention,chan2016listen,chiu2018state} and fully convolutional architectures \cite{zhang2017towards,zeghidour2018fully} are popular while some studies have successfully adopted combination of RNN and convolution neural networks for end-to-end speech recognition \cite{pmlr-v48-amodei16}.
Residual networks \cite{zhang2017very,wang2017residual,kim2017residual}, and highway connections \cite{pundak2017highway} have increased the feasibility of training large, deep neural networks.
A few works have explored joint CTC and sequence-to-sequence learning architectures \cite{watanabe2017hybrid,kim2017joint}.
Self-attention and multi-headed attention based neural networks have yielded state-of-the-art ASR performance \cite{dong2018speech,synnaeve2019end}.

The success of deep neural networks (DNN) is mostly attributed to its ability to utilize vast amounts of data to estimate highly non-linear functions which in turn has resulted in improved acoustic and language modeling.
The lack of suitable available child speech data has limited the modeling capabilities of the DNN models for children speech recognition.
Additionally, the multifaceted signal variability found in children speech poses a number of modeling challenges.
Acoustically, increased speech signal variability in children is mainly attributed to the developmental changes of the vocal apparatus \cite{lee1999acoustics,Lee2014Developmentalacousticstudyof}.
The variability manifests as shifting of formant frequencies, spectral and temporal characteristics both within subject and across subjects and age groups \cite{lee1999acoustics,Lee2014Developmentalacousticstudyof,potamianos2003robust,gerosa2007acoustic}.
Moreover, children speech is characterized with increased pronunciation variability, mispronunciations, disfluencies and non-verbal vocalizations \cite{potamianos1997automatic,potamianos2003robust}. Children's speech is known to include repetitions and revisions \cite{gallagher1977revision}. Also, the use of language, linguistic and grammatical constructs vary with children.
The pronunciation and linguistic variability in children can be attributed to the developing linguistic knowledge and behavior of children.

To address the increased speech signal variability in children, several robust speech signal features, filters and models have been studied and introduced.
Several front-end feature transformations, frequency warping, speaker normalization and filtering techniques have been found to be useful \cite{potamianos2003robust,ghai2009exploring,sinha2018assessment} for mitigating feature space acoustic variability in children.
Vocal Tract Length Normalization (VTLN) \cite{shivakumar2014improving,giuliani2003investigating} techniques have been found particularly beneficial for reducing acoustic variability in robust recognition of children speech.
Adaptation techniques such as maximum likelihood linear regression (MLLR) transforms \cite{shivakumar2014improving,giuliani2006improved,shivakumar2020transfer} also aid in adapting to varying acoustic speech patterns found in children.
Speaker adaptive training based on constrained MLLR \cite{gales1998maximum} was also found to provide notable improvements by reducing heightened inter-speaker variability in children \cite{shivakumar2014improving,giuliani2006improved,shivakumar2020transfer}.
To handle the pronunciation variability, adopting customized dictionaries \cite{li2002analysis} for children and pronunciation modeling techniques \cite{shivakumar2014improving} have been successful.
Finally, to capture linguistic variability and language use of children, language models trained on children's speech have been effective giving improved WER \cite{shivakumar2014improving,shivakumar2020transfer,das1998improvements}.

However, most of the effective modeling adaptations like VTLN, MLLR, speaker adaptive training etc., for children speech are restricted to GMM-HMM and DNN-HMM systems.
This raises questions about the feasibility of newer end-to-end based models for children speech recognition.
Although, few works have explored end-to-end speech recognition for children (in Mandarin language), the benefits and their application to handling various aspects of children speech variability has been unclear \cite{ng2020cuhk,chen2020data,yu2020slt}.
Moreover, these works use fairly limited amount of children speech data (less than 60 hours).

In this paper, we conduct a methodological study into children speech recognition, particularly investigating the most recent developments in end-to-end speech recognition with established state-of-the-art systems. It aims to contribute by answering the following questions:
\begin{enumerate}
\item Do the benefits established with end-to-end speech recognition systems for adult's speech translate to children speech?
\item How do the end-to-end systems compare to the optimized existing DNN-HMM based children speech recognition systems?
\item Will an end-to-end system's ability to exploit large amounts of speech data impute for the anomalies found in children speech?
\item Which neural network based end-to-end architectures are most effective for children speech recognition?
\item How do the end-to-end systems perform for children of different age categories? 
\item What are the merits/demerits of the end-to-end systems compared to DNN-HMM based systems?
\end{enumerate}

The rest of the paper is organized as follows: Section~\ref{sec:acoustic_modeling} presents the different DNN architectures including the DNN-HMM systems and the more recent state-of-the-art end-to-end systems investigated in this study.
Section~\ref{sec:language_modeling} describes the language models (LM) and their architectures employed in our work.
Different decoding techniques investigated as a part of this study are presented in section~\ref{sec:decoding}.
The children speech databases used in this study are listed in section~\ref{sec:databases} and the experimental setup is described in section~\ref{sec:exp_setup}.
Section~\ref{sec:results} presents the results of children speech recognition on adult acoustic models (AM) and the results on children acoustic models are presented in section~\ref{sec:results_child_am}.
More insights on the recognition performance including children age, amount of data, length of utterance, error analysis are carried in section~\ref{sec:error_analysis}.
Finally the study is concluded in section~\ref{sec:conclusion}.

\section{Acoustic Modeling}\label{sec:acoustic_modeling}
In this section, we describe the architectures employed for acoustic modeling in this work.
We select three recently proposed end-to-end architectures that have demonstrated state-of-the-art results in speech recognition on popular benchmarking datasets.
Additionally, for reference to previous works employing DNN-HMM systems, we also consider a competitive DNN-HMM based speech recognition system.
In case of end-to-end architectures, we consider two sets of architectures each, one trained completely supervised on LIBRISPEECH and the other trained on semi-supervised model on LIBRIVOX \cite{synnaeve2019end}.

\subsection{Factorized Time Delay Neural Network (TDNN-F) HMM System}
Factorized Time Delay Neural Networks are one-dimensional convolutional neural networks (CNN) with special semi-orthogonal constraints \cite{povey2018semi}.
The constraints mimic the singular value decomposition in factorizing the weight matrices into products of 2 smaller factors obtained by dropping small singular values.
This enables to preserve the descriptive power of transformations by significantly reducing the number of parameters.
The TDNN-F can be conceptually viewed as introducing an additional bottleneck layer to a traditional convolutional layer (TDNN).
TDNN-F was first introduced for speech recognition giving comparable performance to that of TDNN-LSTM system with almost half the parameters \cite{povey2018semi}.
More recently, TDNN-F models have proven their efficacy for children speech recognition \cite{wu2019advances}.

The architecture made use in our study is comprised of 16 TDNN-F blocks with skip-connections.
Each block consists of a TDNN-F layer followed by rectified linear unit (RELU) non-linearity, followed by a batch normalization layer and a dropout layer.
Each TDNN-F layer has a 1536 dimensional TDNN layer and a 160 dimensional bottleneck layer.
Lattice-free maximum mutual information (LF-MMI) criterion \cite{povey2016purely} is adopted for training the TDNN-F acoustic model.
L2-regularization is adopted during training.
We do not use VTLN since its efficacy in conjunction with TDNN-F was not clear from \cite{wu2019advances}.

\subsection{Residual Neural Network (ResNet)}\label{ssec:resnet}
The ResNet was first proposed for the task of image recognition \cite{he2015deep}.
Increasing the depth of DNN allows for modeling more complex functions, however, the optimization, convergence of the DNNs gets harder as the depth of the network increases and thus limits the number of layers in the network.
This is partly attributed to gradients getting too small (vanishing gradient) or too high (exploding gradients).
The ResNets model the residual functions using skip-connections (shortcut connections skipping a block of layers) rather than the original unreferenced mapping.
It has been found that optimizing the referenced residual functions are easier and alleviate the vanishing/exploding gradient problem, thereby allowing for deeper networks to estimate complex functions efficiently \cite{he2015deep}. 
ResNets have been adopted successfully for speech recognition \cite{xiong2016achieving,saon2017english,wang2017residual}.
Both LSTM \cite{zhang2017very,kim2017residual} and convolution \cite{zhang2017very,xiong2016achieving,saon2017english,wang2017residual} blocks have been proposed with skip connections for ASR.

In this work, we employ the architecture proposed in \cite{synnaeve2019end}.
The input signal is processed using a SpecAugment layer \cite{park2019specaugment} and mapped to an embedding space of dimension 1024 using 1-D convolution layer with stride 2.
The ResNet encoder comprises 12 blocks of 3 1-D convolution layers with a kernel size of 3.
Each convolution layer is followed by ReLU non-linearity, dropout and LayerNorm \cite{ba2016layer}.
The dropout and hidden units increase with depth of the network and additional convolution layers are inserted between ResNet blocks for increasing the hidden dimension.
Three max pooling layers with stride 2 are inserted after block 3, 7 and 10.
The encoder architectures are identical for both CTC and sequence-to-sequence loss, except that the encoder for the sequence-to-sequence has lower dropout for deeper layers and the last bottleneck layer is removed.
The decoder for the sequence-to-sequence model has 2 rounds of key-value attention (see equation~\ref{eq:attn}) as in \cite{hannun2019sequence,vaswani2017attention} through 3 (LIBRISPEECH AM) or 2 (LIBRIVOX AM) layers of RNN-GRU of dimension 512 each followed by a dropout layer.

\subsection{Time-Depth Separable (TDS) Convolution Networks}\label{ssec:tds}
\cite{hannun2019sequence} introduced time-depth separable convolutions for speech recognition with a sequence-to-sequence end-to-end architecture.
The TDS architecture has been shown to generalize better than typical deep convolutional architectures with fewer parameters.
Significant improvements were achieved on the LIBRISPEECH dataset with TDS layers compared to models based on RNN and convolutional networks \cite{hannun2019sequence}.

The core concept of the TDS block is to separate time aggregation from channel mixing and thus increase the receptive field.
The TDS block comprises a 2-D convolutional layer followed by ReLU non-linearity and residual connection followed by LayerNorm \cite{ba2016layer}. 
Finally, the output is re-viewed and is followed by two fully-connected layer with ReLU non-linearity in between and layer normalization.
Moreover, a sub-sampling factor of 8 is applied using 3 sub-sampling layers with stride of 2 each.
The sub-sampling layers are followed with a RELU and layer normalization.

The architecture used in this study is similar to \cite{synnaeve2019end}.
The input signal is processed using a SpecAugment layer \cite{park2019specaugment} and mapped to an embedding space using 2D-convolution layer with stride of size 2$\times$1.
For LIBRISPEECH AM, 3 groups of TDS blocks are employed, containing 5, 6 and 10 TDS blocks each with 10, 14, and 18 channels respectively.
For LIBRIVOX AM, 4 groups of TDS blocks are employed, containing 2, 2, 5 and 6 TDS blocks each with 16, 16, 32, and 48 channels respectively.
The number of channels in the feature maps spanning the two internal fully-connected layers are increased by a factor of 3 (LIBRISPEECH AM), or 2 (LIBRIVOX AM) via sub-sampling 2D-convolutional layers.
All the 2D-convolutional layers are followed by ReLU non-linearity and LayerNorm \cite{ba2016layer}.
The kernel size of both the TDS blocks and 2D-convolutions is set to 21$\times$1 (LIBRISPEECH AM), or 21$\times$3 (LIBRIVOX AM).
The encoder architecture is identical for both CTC and sequence-to-sequence models.
In-case of sequence-to-sequence networks, the decoder architecture is identical to the ResNet decoder network described in section~\ref{ssec:resnet}, i.e., the decoder network has 2 rounds of key-value attention (see equation~\ref{eq:attn}) through 3 (LIBRISPEECH AM) or 2 (LIBRIVOX AM) layers of RNN-GRU of dimension 512 each followed by a dropout layer.

\subsection{Transformers}\label{ssec:transformer}
Transformer networks were first introduced for the task of machine translation \cite{vaswani2017attention} significantly advancing the state-of-the-art.
Since then, transformers have dominated the fields of natural language processing \cite{devlin2018bert}, speech recognition \cite{dong2018speech}, spoken language technologies \cite{karita2019comparative} as well as the computer vision and image processing domains \cite{parmar2018image}.
The transformer is a neural sequence transducer with an encoder-decoder architecture based solely on attention mechanisms.
They employ 6 stacked multi-headed self-attention layers each followed by fully connected layers for both encoder and decoder.
The self-attention is described in terms of mapping a query and a set of key-value pairs to an output.
The self-attention is defined as:
\begin{equation}\label{eq:attn}
Attention(Q,K,V) = softmax(\frac{QK^{T}}{\sqrt{d_{k}}})V
\end{equation}
It is essentially softmax weighted sum of values, $V$, where the weights are dot-product of two matrices $Q$ (query) and $K$ (keys) each corresponding to collection of sequence of input vectors which are scaled by the dimension, $d_k$, of the key vectors.
The term multi-head refers to projecting the key, value and query vectors into multiple subspaces and running multiple self-attention in parallel on each to derive multiple outputs and concatenating the outputs.
The multi-head attention is given by:
\begin{align}
\text{MultiHead}&(Q,K,V) = \text{Concat}(\text{head}_1,\ldots,\text{head}_n)W^M \nonumber \\
\text{head}_i &= Attention(QW^{Q}_{i},KW^{K}_i,VW^{V}_{i}) 
\end{align}
where $W^Q_i$, $W^K_i$, $W^V_i$ are projections corresponding to head $i$, for query, key and value respectively.

In this work, we adopt the architecture specified in \cite{synnaeve2019end} for training the acoustic model.
A front-end of 3 (LIBRISPEECH AM) or 6 (LIBRIVOX AM) layers of 1-D convolutions each with kernel size 3, strided by 8 frames (80ms) is used as feature extraction for the subsequent transformer blocks. 
The (input, output) size of the first layer is $(80, D_c)$, the last layer is $(D_c/2,D_{tr}\times 2)$ and the intermediate layers is $(D_c/2,D_c)$, with $D_c$=1024, $D_{tr}$=1024 for self-attention, 4096 for feed-forward network (LIBRISPEECH AM) or $D_c$=2048, $D_{tr}$=768 for self-attention, 3072 for feed-forward network (LIBRIVOX AM).
Each convolution layer is strided by 2 each (LIBRISPEECH AM) or by 2 every alternate layer (LIBRIVOX AM). 
Each convolution is followed by gated linear unit (GLU), dropout and LayerNorm.
Next, with each succeeding transformer block 4-head attention is used with skip (residual) connection followed by layer normalization, feed-forward layer and one hidden layer with RELU non-linearity.
Additionally, skip (residual) connection is used across the entire transformer block.
Dropouts are used on the self-attention.
Moreover layer-drop \cite{fan2019reducing} is employed for feed-forward network to drop the entire layer.
The encoder consists of 24 (LIBRISPEECH AM), or 36 (LIBRIVOX AM) stacked transformer blocks.
Identical architecture is used for encoder of both CTC and sequence-to-sequence models.
For sequence-to-sequence models, the decoder is made up of 6 stacked Transformers with 4 attention heads and encoding dimension of 256.

\section{Language Modeling}
\label{sec:language_modeling}
In this section, the language models used in beam-search decoding are described.
We experiment with 4 types of language models: (i) word-based 4-gram LM, (ii) word-piece 6-gram LM, (iii) word-based gated convolutional neural network (GCNN) LM, and (iv) word-piece based GCNN LM \cite{dauphin2017language}.
Since we employ a lexicon during decoding for CTC based models, word-based 4-gram LM amd GCNN LM are restricted to CTC models.
In case of sequence-to-sequence models, we employ lexicon-free decoding \cite{likhomanenko2019needs} along with word-piece based 6-gram LM and GCNN word-piece LM.

Gated convolutional neural networks were first proposed for the task of language modeling \cite{dauphin2017language}.
GCNN is among the first non-recurrent, highly parallelizable, finite context approach with stacked convolutions to give competitive, low latency alternative to strong recurrent language models.
The gating mechanism alleviates the vanishing gradient problem enabling deeper networks for language modeling.
The gating operation in GCNN is formulated as:
\begin{equation}
h(X) = (X*W+b) \otimes \sigma (X*V+c)
\end{equation}
where $h$ is the hidden layer, $X\in \mathbb{R}^{N\times m}$ is the input for layer $h$, $W$ and $V$ are the weights of the 1-D convolution layer $\in \mathbb{R}^{k\times m \times n}$, $b$ and $c$ are the biases $\in \mathbb{R}^n$, $\sigma$ is the sigmoid function, $\otimes$ denotes element wise product and m, n, k and N are the input feature map, output feature map, patch size and length of the input sequence respectively.

The architectures of the GCNN LM are borrowed from \cite{dauphin2017language} (GCNN-14B architecture).
The GCNN-14B bottleneck architecture comprises of an embedding layer which maps the input words to a fixed dimension of 1024.
The embedding layer is followed by 14 residual blocks.
The first residual block contains one gated 1-D convolution layer with [kernel size, output size] of [5, 512].
Residual blocks 2 to 4 are comprised of 3 gated 1-D convolution layer each with [1,128], [5,128], [1,512]; 
residual blocks 5 to 7 comprised of 3 gated 1-D convolution layer each with [1,512], [5,512], [1,1024]; 
residual blocks 8 to 13 comprised of 3 gated 1-D convolution layer each with [1,1024], [5,1024], [1,2048] and
the final residual block contains one gated 1-D convolution layer with [1,1024], [5,1024], [1,4096].
The softmax layer outputs the probability distribution over all the words/word-pieces in the vocabulary.

\begin{table*}[t]
\centering
\begin{tabular*}{0.8\textwidth}{l@{\extracolsep{\fill}}lll}
\toprule
Corpus & Train & Development & Test \\
\midrule
\multirow{3}{*}{MyST} & 88318 Utterances & 5000 Utterances & 5000 Utterances \\
& 197.72 hours & 12.23 hours & 13.28 hours \\
& 678 speakers & 25 speakers & 34 speakers \\
\midrule
\multirow{3}{*}{OGI Kids} & & & 1099 Utterances \\
& & & 30.5 hours \\
& & & 1099 speakers \\
\bottomrule
\end{tabular*}
\caption{Statistics: Children speech corpus}\label{tab:database}
\end{table*}

\section{ASR Decoding}
\label{sec:decoding}
Decoding is the process of scoring the hypothesis with the acoustic model and the language model to derive the final output.
In this study, we assess two specific types of decoding (i) beam-search decoder, and (ii) greedy decoder.

\subsection{Beam-search Decoder}
\label{ssec:beamsearch}
The output of the neural networks can be viewed as a $C \times T$ matrix (lattice) with probabilities over each class $c\in \{1\ldots C\}$ for each time step $t\in \{1\ldots T\}$.
Each path through the lattice represents a possible ASR hypothesis which can be scored by a LM to further influence the scores of acoustic model.
A typical beam-search decoder outputs a hypothesis that maximizes:
\begin{equation}
log P_{AM}(y|x) + \alpha log P_{LM}(y) + \beta |y|
\end{equation}
where $y$ is the output hypothesis, $x$ is the input acoustic features, $\alpha$ is the LM weight and $\beta$ is the word insertion penalty.
Additionally, for sequence-to-sequence models end-of-sentence (EOS) penalty is adopted to control the output hypothesis lengths.
The LM weight, word insertion penalty and EOS penalty are all tuned using grid search in our experiments.
In order to keep the memory and computation complexity tractable, top few states are considered over each time-step.
The number of top states considered defines the beam-size.

In our experiments, we consider two types of beam-search decoding, (i) lexicon-based, and (ii) lexicon-free.
With lexicon-based decoding, a dictionary mapping is used to convert the output of the acoustic model to words, and thus the beam-search space is restricted to words in the dictionary.
Whereas, with lexicon-free decoding, the beam-search space is not restricted to words and operates on word-pieces, thus capable of outputting words with arbitrary spellings.
The lexicon-based decoding requires the LM to be on word-level and the lexicon-free decoding requires the LM with input tokens as word-pieces.
As suggested in \cite{synnaeve2019end}, we adopt lexicon-based decoding for models trained with CTC loss and lexicon-free decoding for sequence-to-sequence models.

\subsection{Greedy Decoder}
\label{ssec:greedy}
With greedy decoding, there is no language model involved, and the most probable output of the acoustic model is considered as final.
The end-to-end acoustic models are capable of learning language model inherently given enough training data.
Studies such as \cite{synnaeve2019end}, have shown that given large amounts of data, the greedy decoding without language model performs just as good as beam-search decoding with a large language model \cite{synnaeve2019end}.

\section{Databases}
\label{sec:databases}

The choice of the children's speech corpora in our study is mainly based on (i) amount of children speech available, and (ii) good distribution of age demographics among the children for analysis purposes.
We make use of two popular children's speech corpora:

\subsection{My Science Tutor (MyST) Children Speech Corpus}
The MyST Corpus \cite{ward2011my,ward2019my} is one of the publicly available large collection of English children's speech.
It consists of 499 hours with 244,069 utterances of conversational speech between children and a virtual tutor.
The corpus consists of 1,372 students from third, fourth and fifth grades having conversations spanning 9 areas of science.
This makes the corpora larger than all other available children's English speech corpora combined together.
However, only 42\% of the corpus is annotated for ASR purposes, i.e., 103,429 utterances (233 hours).
The transcribed subset of the corpora were further cleaned and 98,318 utterances (223.23 hours) with 737 speakers were retained.
The database is randomly split into three parts for training, development and held-out test set without speaker overlap.
The details of the split is presented in Table~\ref{tab:database}.

\subsection{OGI Kids Speech Corpus}
\begin{figure}[t]
\includegraphics[width=\columnwidth]{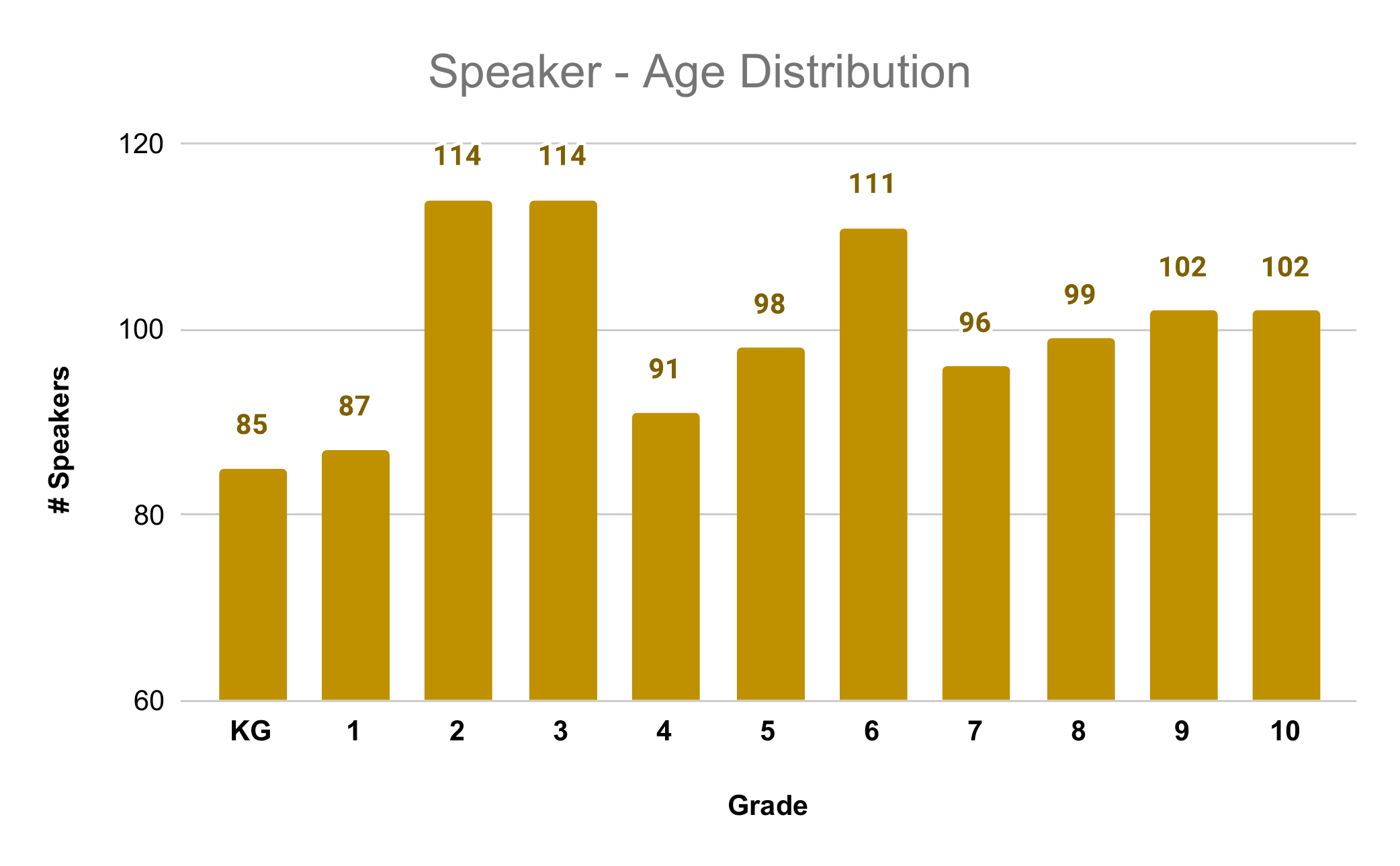}
\caption{Speaker Age Distribution for OGI Kids Corpus\\
\small{KG refers to Kindergarten}}\label{fig:database}
\end{figure}

To have a broad range of age demographics among children, for investigating age related effects, we additionally make use of the OGI Kids speech corpus \cite{shobaki2000ogi}.
The OGI Kids corpus consists of 1100 children ranging from kindergarten to 10th grade.
In this study, we only select the spontaneous speech subset of the data with annotated transcripts since the spontaneous children speech is believed to be more complex both in acoustic and linguistic constructs compared to the prompted speech \cite{gerosa2006acoustic}.
In the spontaneous speech data portion, the experimenter asks the child a series of questions to elicit a spontaneous response.
In our study, we use this corpus for evaluation purposes only. 
The statistics are presented in Table~\ref{tab:database} and the age distributions are presented in Figure~\ref{fig:database}.

\section{Experimental Setup}
\label{sec:exp_setup}
\subsection{Hybrid TDNN-F HMM Acoustic Model}
The Kaldi ASR toolkit \cite{povey2011kaldi} was used for training the TDNN-F based hybrid DNN-HMM acoustic model.
For the baseline adult models, we use the pre-trained models trained on LIBRISPEECH\footnote{\url{https://www.openslr.org/11/}} \cite{panayotov2015librispeech} made available by KALDI developers\footnote{\url{http://kaldi-asr.org/models/m13}}.
13-dimensional Mel-filter cepstral coefficients (MFCC) features were extracted with a window size of 25ms and window shift of 10ms with delta and delta-delta coefficients for training GMM-HMM system.
A GMM-HMM system with linear discriminant analysis (LDA), maximum likelihood linear transform (MLLT) and feature-space maximum likelihood linear regression (fMLLR) based speaker adaptive training (SAT) is used to obtain the alignments needed to train the TDNN-F acoustic model. 
40-dimensional MFCC features were used with left and right context of 1 frame along with 100 dimensional i-vector features to train the TDNN-F acoustic model.
The i-vectors were trained in-domain on LIBRISPEECH using 40-dimensional MFCC features with left and right context of 3 and a subsequent PCA dimension reduction.
The TDNN-F acoustic model is trained to predict among 6024 Gaussian mixtures.

\subsection{Hybrid TDNN-F HMM Acoustic Model for Children}
For adaptation on children data, we perform transfer learning due to its performance advantages on children's speech data \cite{shivakumar2020transfer}.
We initialize the acoustic model with the pre-trained adult model trained on LIBRISPEECH.
The last layer is removed and a new randomly initialized TDNN-F and output linear layers are added to the model.
The transferred layers are updated with a smaller learning rate (0.25 times) while the newly added layers are trained with a higher learning rate on MyST training corpus.
The MyST corpus is forced aligned using the pre-trained model and the alignments are obtained.
The i-vectors for children data are extracted from the LIBRISPEECH i-vector model.
The TDNN-F model is optimized for LF-MMI criterion using stochastic gradient descent with 0.001 learning rate trained for 4 epochs.
The convergence is ensured using the development subset of the MyST corpus.

\subsection{End-to-end Acoustic Model}
All the end-to-end ASR experiments are carried out with the wav2letter++ toolkit \cite{pratap2018wav2letter++}.
For evaluations on the baseline (un-adapted) adult models we use two versions of models presented in \cite{synnaeve2019end}: (i) model trained on LIBRISPEECH \cite{panayotov2015librispeech}, and (ii) semi-supervised model trained on LIBRIVOX\footnote{\url{https://librivox.org}}.
The model trained on LIBRISPEECH is fully supervised.
The supervised LIBRISPEECH model is further used to decode the entire LIBRIVOX database to generate the labels for the unlabeled LIBRIVOX dataset.
For this purpose, a Transformer network trained with CTC loss with beamsearch decoding using 4-gram language model is employed.
The semi-supervised model is trained by combining the LIBRISPEECH with true labels along with the labels generated for the LIBRIVOX corpora.
Since the semi-supervised LIBRIVOX model has data orders of magnitude more than LIBRISPEECH, two set of architectures are used differing in the number of parameters.
In this paper, we utilize the pre-trained acoustic models open sourced\footnote{\url{https://github.com/facebookresearch/wav2letter/tree/v0.2/recipes/models/sota/2019}} for adult ASR.
More details regarding the experimental setup and hyper-parametrization of the models can be found in \cite{synnaeve2019end}.

\subsection{End-to-end Acoustic Model for Children}
For adaptation with children's speech data, instead of training model from scratch, we initialize the acoustic model with the adult models trained on LIBRISPEECH and LIBRIVOX and fine-tune the entire model as suggested in \cite{shivakumar2020transfer}.
The end-to-end AMs are trained using 80-dimensional (channel) log-mel filterbank features extracted using Hamming window with window shift of 10ms and a window size of 25ms for Transformers and window size of 30ms for TDS and ResNet models.
All the acoustic models output probability distribution over 10k word pieces \cite{kudo2018sentencepiece} generated using the SentencePiece toolkit\footnote{\url{https://github.com/google/sentencepiece}}.

For ResNet and TDS based models, the batch-size is set to 4, the dropout is in the range [0.05, 0.2] increasing with depth.
The momentum is set to 0.5 for ResNet-CTC, 0.1 for ResNet-S2S, 0.1 for TDS-CTC and 0.0 for TDS-S2S model.
In the case of Transformer models, linear learning rate warm-up is applied for 30k updates, the dropout and layer-drop is set to 0.2 for all Transformer blocks, the momentum is set to 0.95, batch size of 5 is adopted.
For training sequence-to-sequence models, 99\% teacher forcing, 1\% word-piece sampling, 5\% label smoothing is employed.
For sequence-to-sequence Transformer models, dropout and layer-drop in the decoder is set to 0.1.
The learning rate is set to 0.01 with a step-wise learning rate schedule decreasing by a factor of 2 every 150 updates.
Stochastic Gradient Descent (SGD) is used for updating ResNet, TDS models and Adagrad is used for the Transformers.
The models are fine-tuned for 10 epochs and convergence is ensured.

During beamsearch decoding, we use a beam-size of 500 for CTC models and 50 for sequence-to-sequence models, the LM weights are tuned in [0.1, 1.3], word insertion penalty in the range [0.1, 1.3] and the EOS penalty in the range [-10.0, -4.0] on the development dataset.

\subsection{Language Models}
The base language models are trained on the LIBRISPEECH LM corpus\footnote{\url{https://openslr.org/11/}} containing data from 14,500 public domain books.
The 4-gram word LM and 6-gram word-piece LM are trained using the KenLM toolkit \cite{heafield2011kenlm}.
The 4-gram LM does not employ any pruning, however the word-piece based 6-gram models involve pruning 5-grams once and 6-gram appearing twice or fewer.
The GCNN LMs are trained using the fairseq toolkit\footnote{\url{https://github.com/pytorch/fairseq}} \cite{ott2019fairseq}.
More details regarding the setup can be found in \cite{likhomanenko2019needs} and \cite{synnaeve2019end}.

For including children's data for language modeling, we make use of the text from MyST training subset of the corpus.
In case of n-gram based models, independent LMs are trained, i.e., one word-based 4-gram and one word-piece based 6-gram model with similar setup as described earlier.
Next, the children LM is interpolated with the LIBRISPEECH LM by tuning weights on the development set of MyST corpus text data.
In case of GCNN LM, the neural network is initialized with the weights from corresponding LIBRISPEECH LMs and then fine-tuned with the MyST train subset.
The GCNN is optimized with Nesterov accelerated gradient descent for 20 epochs with a learning rate of 0.0001 and momentum of 0.99.
Gradient clipping and weight normalization are employed for stabilization \cite{dauphin2017language}.

\begin{table*}[t]
\centering
\begin{tabular}{lllcccccc}
\multirow{2}{*}{} & \multirow{2}{*}[-1em]{AM} & \multirow{2}{*}[-1em]{LM} & \multicolumn{2}{c}{LIB test-clean} & \multicolumn{2}{c}{MyST test} & \multicolumn{2}{c}{OGI Kids} \\
\cmidrule(lr){4-5} \cmidrule(lr){6-7} \cmidrule(lr){8-9} \\
& & & LER & WER & LER & WER & LER & WER \\
\midrule
KALDI & TDNN-F DNN-HMM & 4-gram & 2.22 & 5.94 & 26.98 & 47.90 & 36.04 & 53.55 \\
\midrule
\multirow{6}{*}{Greedy Decoding} & ResNet + CTC & - & 1.57 & 4.25 & 21.24 & 36.82 & \textbf{33.42} & 52.06 \\
& ResNet + S2S & - & 2.45 & 4.92 & 38.19 & 54.37 & 75.19 & 86.32 \\
& TDS + CTC & - & 1.85 & 4.80 & 25.00 & 41.20 & 36.49 & 54.74 \\
& TDS + S2S & - & 1.38 & 3.43 & 27.70 & 42.13 & 77.95 & 84.74 \\
& Transformer + CTC & - & 1.14 & 3.29 & \textbf{16.86} & 29.25 & 40.58 & 50.37 \\
& Transformer + S2S & - & 1.02 & 2.89 & 25.88 & 38.81 & 74.14 & 87.22 \\
\arrayrulecolor{black!50}\midrule
\multirow{12}{*}{Beamsearch Decoding} & ResNet + CTC & 4-gram & 1.53 & 3.68 & 20.78 & 33.97 & 33.56 & 49.19 \\
& ResNet + S2S & 6-gram-wp & 1.72 & 3.88 & 56.54 & 76.53 & 82.85 & 92.40 \\
& TDS + CTC & 4-gram & 1.77 & 3.98 & 25.10 & 38.03 & 37.40 & 52.32 \\
& TDS + S2S & 6-gram-wp & 1.28 & 3.18 & 32.40 & 47.15 & 87.70 & 89.78 \\
& Transformer + CTC & 4-gram & 1.19 & 2.88 & 17.84 & 27.54 & 51.76 & 55.69 \\
& Transformer + S2S & 6-gram-wp & 1.06 & 2.72 & 37.46 & 51.54 & 72.78 & 88.88 \\
\arrayrulecolor{black!20}\cmidrule(lr){2-9}
& ResNet + CTC & GCNN & 1.45 & 3.28 & 20.49 & 32.48 & 34.27 & \textbf{49.06} \\
& ResNet + S2S & GCNN-wp & 1.85 & 3.79 & 64.98 & 86.09 & 83.13 & 94.36 \\
& TDS + CTC & GCNN & 1.63 & 3.40 & 25.73 & 36.28 & 39.59 & 52.15 \\
& TDS + S2S & GCNN-wp & 1.17 & 2.93 & 38.33 & 53.77 & 87.58 & 90.26 \\
& Transformer + CTC & GCNN & 1.12 & 2.58 & 17.43 & \textbf{26.23} & 52.58 & 55.92 \\
& Transformer + S2S & GCNN-wp & \textbf{0.90} & \textbf{2.40} & 32.73 & 45.96 & 72.27 & 88.84 \\
\arrayrulecolor{black}\bottomrule
\end{tabular}
\caption{Results on models trained on LIBRISPEECH}\label{tab:res_librispeech}
\end{table*}

\section{Results: Adult Acoustic Models}
\label{sec:results}

In this section, we present the results comparing the DNN-HMM and the state-of-the-art end-to-end acoustic models trained on adult speech for application to children speech recognition.

\subsection{Adult Speech Recognition}
Table~\ref{tab:res_librispeech} lists the DNN-HMM system and various end-to-end ASR system both trained on exactly same data (960 hours of LIBRISPEECH) and incorporates identical language models.
It is observed that testing on test-clean subset of LIBRISPEECH adult speech, the TDNN-F based DNN-HMM system achieves a WER of 5.94\%. Comparatively, the best performing end-to-end ASR based on Transformer architecture with sequence-to-sequence training incorporating gated-CNN (GCNN) word-piece language model achieves a WER of 2.4\%, i.e., a relative improvement of 59.6\%.
In terms of LER, the relative improvement is similar i.e., 59.46\%.

\begin{table*}[t]
\centering
\begin{tabular}{lllcccccc}
\multirow{2}{*}{} & \multirow{2}{*}[-1em]{AM} & \multirow{2}{*}[-1em]{LM} & \multicolumn{2}{c}{LIB test-clean} & \multicolumn{2}{c}{MyST test} & \multicolumn{2}{c}{OGI Kids} \\
\cmidrule(lr){4-5} \cmidrule(lr){6-7} \cmidrule(lr){8-9} \\
& & & LER & WER & LER & WER & LER & WER \\
\midrule
\multirow{6}{*}{Greedy Decoding} & ResNet + CTC & - & 0.93 & 2.74 & 16.81 & 28.26 & 25.75 & 38.00 \\
& ResNet + S2S & - & 1.11 & 2.70 & 28.11 & 41.07 & 68.33 & 79.77 \\
& TDS + CTC & - & 0.98 & 2.85 & 17.71 & 29.25 & 26.11 & 38.24 \\
& TDS + S2S & - & 0.85 & 2.40 & 21.06 & 32.29 & 73.48 & 76.49 \\
& Transformer + CTC & - & 0.87 & 2.59 & \textbf{15.71} & 25.46 & 47.42 & 54.33 \\
& Transformer + S2S & - & \textbf{0.76} & 2.28 & 18.78 & 29.01 & 80.44 & 85.32 \\
\arrayrulecolor{black!50}\midrule
\multirow{18}{*}{Beamsearch Decoding} & ResNet + CTC & 4-gram & 1.04 & 3.88 & 16.59 & 28.89 & \textbf{25.02} & 37.32 \\
& ResNet + S2S & 6-gram-wp & 1.10 & 2.66 & 26.21 & 36.65 & 77.53 & 84.54  \\
& TDS + CTC & 4-gram & 1.12 & 2.79 & 18.17 & 28.15 & 27.59 & \textbf{37.18} \\
& TDS + S2S & 6-gram-wp & 0.86 & 2.40 & 21.05 & 31.93 & 71.94 & 74.62 \\
& Transformer + CTC & 4-gram & 1.04 & 2.52 & 17.57 & 25.21 & 54.70 & 58.66 \\
& Transformer + S2S & 6-gram-wp & 0.79 & 2.25 & 25.91 & 39.25 & 67.85 & 81.56 \\
\arrayrulecolor{black!20}\cmidrule(lr){2-9}
& ResNet + CTC & GCNN & 1.09 & 2.45 & 18.33 & 26.00 & 31.59 & 37.78 \\
& ResNet + S2S & GCNN-wp & 1.10 & 2.65 & 27.43 & 37.77 & 86.36 & 90.64 \\
& TDS + CTC & GCNN & 1.16 & 2.54 & 19.28 & 27.01 & 30.77 & 37.42 \\
& TDS + S2S & GCNN-wp & 0.86 & 2.27 & 23.93 & 36.22 & 70.48 & 77.32 \\
& Transformer + CTC & GCNN & 1.03 & 2.41 & 16.79 & \textbf{24.26} & 52.15 & 55.67 \\
& Transformer + S2S & GCNN-wp & 0.80 & \textbf{2.17} & 26.89 & 40.51 & 70.77 & 85.15 \\
\arrayrulecolor{black!20}\cmidrule(lr){2-9}
& ResNet + CTC & 4-gram (M) & - & - & 17.05 & 24.92 & 30.08 & 37.31 \\
& ResNet + S2S & 6-gram-wp (M) & - & - & 30.01 & 41.27 & 73.12 & 83.91 \\
& TDS + CTC & 4-gram (M) & - & - & 17.95 & 25.92 & 30.41 & 39.80 \\
& TDS + S2S & 6-gram-wp (M) & - & - & 21.07 & 30.82 & 74.91 & 76.46 \\
& Transformer + CTC & 4-gram (M) & - & - & 16.47 & \textbf{23.32} & 53.50 & 56.50 \\
& Transformer + S2S & 6-gram-wp (M) & - & - & 19.48 & 27.81 & 86.17 & 88.32 \\
\arrayrulecolor{black}\bottomrule
\end{tabular}
\caption{Results on models trained on LIBRISPEECH + LIBRIVOX(58k hours)\\
\small{(M) refers to LM interpolated with MyST model}}\label{tab:res_librivox}
\end{table*}

\subsection{Children Speech Recognition}
Columns 6-9 of Table~\ref{tab:res_librispeech} list the results of Children's speech recognition on MyST Kids corpus and the OGI Kids Corpus.
First, we observe that both the LER and WER increases for children speech, and the results for OGI Kids corpus is relatively worse compared to MyST Kids Corpus.
This is expected since the MyST Corpus contains speech data for children in 3-5 grades, whereas the OGI Kids corpus contains children ranging from Kindergarten to 10th grade (see Figure~\ref{fig:database}).
We believe the inclusion of data for younger children i.e, kindergarten to 3rd grade in OGI Kids Corpus is the main factor for lower performance compared to MyST Corpus.
Assessing the improvements with the end-to-end ASR over the DNN-HMM system, while modest improvements of relative 45.24\% reduction in WER (37.51\% reduction in LER) is observed in case of MyST corpus, only 8.38\% reduction in WER (7.27\% reduction in LER) is observed with OGI Kids corpus.

In comparison to the corresponding adult acoustic models, for MyST Corpus with the TDNN-F HMM system, the WER is over 7 times worse for children and for the best performing end-to-end based ASR the WER is nearly 10 times worse.
For the OGI Kids Corpus, with the TDNN-F DNN-HMM system, the WER is over 8 times worse for children and for the end-to-end ASR the WER is nearly 19.5 times worse in comparison to adult speech recognition.
Although the end-to-end systems give improvements in absolute WERs compared to the DNN-HMM based systems, they undergo a higher degree of degradation and are relatively less generalizable towards children speech.
Overall, the state-of-the-art end-to-end systems setting high benchmarks on adult speech are far from achieving the same level of performance for children speech.

\subsection{Effect of amount of training data for Acoustic Models}
In this section, we assess the results of exploiting large amount of adult speech data for training end-to-end acoustic models.
Table~\ref{tab:res_librivox} presents the results with acoustic models trained on combination of LIBRISPEECH (960 hours) and LIBRIVOX (53,800 hours) (semi-supervised). 
Compared to results in Table~\ref{tab:res_librispeech}, the performance on adult's speech (test-clean subset of LIBRISPEECH) improves by relative 22.22\% LER and 9.58\% WER.
Evaluating on children's speech, MyST Corpus, the relative improvements with additional 53,800 hours of training data is 6.82\% LER and 3.89\% WER, and on OGI Kids corpus, the relative improvements is 25.13\% LER and 24.22\% WER.
With our experiments, we find that exploiting large amounts of speech data for acoustic model even with adult's speech, improvements are observed for children's speech recognition.
A detailed analysis on these improvements are provided in section~\ref{sec:error_analysis}.

\subsection{Greedy Decoding versus Beamsearch Decoding}
For adult speech recognition, the best results both with LER and WER are observed with Beamsearch decoding (see Table~\ref{tab:res_librispeech}).
The relative improvement obtained with beamsearch decoding over greedy decoding is 16.96\% WER and 11.76\% in LER.
The beamsearch decoding is able to exploit additional knowledge from language models especially with GCNN based LM to provide considerable improvements over greedy decoding.
However, with significant increase in the training data, see Table~\ref{tab:res_librivox}, evaluating on adults speech, the greedy decoding outperforms the beamsearch decoding in terms of LER (3.8\% reduction) and the gains with beamsearch decoding in terms of WER reduces to 4.82\%.
Overall, with large amount of training data the greedy decoding benefits, and approaches performance of beamsearch decoding by learning an implicit language model \cite{synnaeve2019end}.

For children speech recognition, greedy decoding results in better LER over beamsearch decoding, i.e., 3.27\% for MyST corpus and 0.42\% for OGI Kids (see Table~\ref{tab:res_librispeech}).
However, better WERs are obtained with beamsearch decoding, a relative improvement of 10.32\% for MyST and 2.6\% for OGI Kids.
The greedy decoding benefits more with additional speech data, see Table~\ref{tab:res_librivox}, improvements in order of 30.86\% reduction in WER with MyST corpus and 24.56\% reduction with OGI Kids corpus.
With large data, the performance of greedy decoding approaches that of beamsearch decoding even for children speech, similar to observations made with adult speech recognition \cite{synnaeve2019end}.

\subsection{End-to-End Architectures}\label{ssec:e2e_arch}
For evaluations on adult speech models trained on LIBRISPEECH (Table~\ref{tab:res_librispeech}), we observe that Transformer based architecture gives the best results (18.09\% WER reduction over the TDS networks) followed by the Time-Depth Separable networks and then the Residual Networks.
We find the Transformer based architecture consistently gives better results both in terms of LER and WER for adult speech with both greedy and beamsearch decoding.
This trend also translates to models trained on 54,760 hours of LIBRISPEECH combined with LIBRIVOX, i.e., the Transformer networks give a relative 4.41\% WER reduction over the TDS networks.

With the experiments on children speech, again the Transformer based architecture proves favorable while evaluating on MyST children speech.
The improvements over the ResNet architecture is 19.24\%  (Table~\ref{tab:res_librispeech}).
The addition of training data (54.8k hours) leads to improvement of 19.95\% with ResNets and 25.55\% with TDS networks.
However, the improvements are minimal with Transformers (7.51\%) and the performance advantage of Transformers over ResNets decreases to 6.69\%.

The evaluations on OGI Kids corpus show the ResNets and TDS network outperform the Transformer networks.
Performance of the Transformer networks drops significantly relative to the best results obtained with ResNets and TDS networks for models trained on 960 hours (2.69\% increase in WER).
Addition of training data (54.8k hours) leads to improvements with ResNets (19.95\%) and TDS networks (25.55\%).
But interestingly, the performance of Transformer networks drops by 7.88\% WER.
Overall, the WER with Transformers is 46.13\% worse relative to best performance obtained with TDS network.
We believe the increased variability due to diverse age range in OGI Corpus (inter-age acoustic variability in children) impacts the Transformer networks negatively.
This indicates that the Transformer networks are less generalizable for children speech.
Further analysis on this aspect is provided in section~\ref{sec:error_analysis}.

\begin{table*}[t]
\centering
\begin{tabular}{lllcccc}
\multirow{2}{*}{} & \multirow{2}{*}[-1em]{AM} & \multirow{2}{*}[-1em]{LM} & \multicolumn{2}{c}{MyST test} & \multicolumn{2}{c}{OGI Kids} \\
\cmidrule(lr){4-5} \cmidrule(lr){6-7} \\
& & & LER & WER & LER & WER \\
\midrule
KALDI & TDNN-F DNN-HMM & 4-gram & 11.67 & 19.51 & 30.40 & 44.74 \\
\midrule
\multirow{6}{*}{Greedy Decoding} & ResNet + CTC & - & 12.08 & 19.53 & 22.44 & 35.82 \\
& ResNet + S2S & - & 20.44 & 27.53 & 65.66 & 76.49 \\
& TDS + CTC & - & 11.70 & 20.04 & \textbf{22.19} & 34.56 \\
& TDS + S2S & - & 12.95 & 18.72 & 64.76 & 69.36 \\
& Transformer + CTC & - & \textbf{9.17} & \textbf{16.01} & 36.52 & 49.80 \\
& Transformer + S2S & - & 11.67 & 16.69 & 70.42 & 77.12 \\
\arrayrulecolor{black!50}\midrule
\multirow{6}{*}{Beamsearch Decoding} & ResNet + CTC & 4-gram (M) & 12.48 & 18.23 & 23.84 & 34.73 \\
& ResNet + S2S & 6-gram-wp (M) & 20.27 & 27.03 & 63.04 & 73.85 \\
& TDS + CTC & 4-gram (M) & 11.89 & 18.61 & 23.76 & \textbf{33.64} \\
& TDS + S2S & 6-gram-wp (M) & 13.24 & 18.77 & 60.15 & 65.09 \\
& Transformer + CTC & 4-gram (M) & 10.19 & 16.74 & 37.18 & 48.90 \\
& Transformer + S2S & 6-gram-wp (M) & 11.78 & 16.81 & 63.26 & 71.11 \\
\arrayrulecolor{black}\bottomrule
\end{tabular}
\caption{Results on models fine-tuned on MyST Corpus\\
\small{(M) refers to LM interpolated with MyST model}}\label{tab:res_myst}
\end{table*}

\subsection{CTC versus Sequence-to-Sequence training}\label{ssec:ctc_vs_s2s}
Observations made on test-clean subset of Librispeech indicate that sequence to sequence training gives the best performance for adult speech.
However, the observations are reversed for children speech recognition both with MyST and OGI Kids corpus.
The performance of sequence-to-sequence models are always much worse compared to the CTC counterparts. 
Moreover, the performance of sequence-to-sequence models almost breaks down on the OGI Kids corpus.
We believe this is because the heightened variability found in children in terms of speaking rate, varying phoneme duration and acoustic characteristics poses problem for alignment in case of sequence-to-sequence models which implicitly estimate attention based alignments.
On the other hand, the CTC models with explicit alignments are more robust to children speech.
Another important factor is the utterance lengths for the OGI Kids corpus are much longer than that of MyST corpus, the sequence-to-sequence networks have been shown to have problems with processing long time sequences \cite{chorowski2015attention}.
Another notable observation in our experiments is that in most of the cases the ResNet-CTC models perform better than the TDS-CTC models, while TDS-S2S models perform better than ResNet-S2S models.

\subsection{Language Models}
GCNN based LM provides modest gains over the n-gram models on adult speech recognition.
The gains are more prevalent on models trained on LIBRISPEECH data (relative improvement of 11.76\% WER) versus on acoustic models trained on additional 53,800 hours of LIBRIVOX data (relative improvement of 3.56\% WER).
Decoding on MyST children corpus, we find GCNN LM provides improvements up-to 4.76\% WER on LIBRISPEECH acoustic models, which reduces to improvements of 3.77\% WER with added LIBRIVOX data.
With the OGI Kids corpus, we find GCNN LM to be effective only on LIBRISPEECH acoustic models and they fail to provide improvements on acoustic model trained on additional LIBRIVOX dataset.

Table~\ref{tab:res_librivox} also presents the results of adult acoustic models in conjunction with children LM.
The children LM is a mixture of the LIBRISPEECH LM interpolated along with LM trained on test-corpus of MyST Corpus.
The results show definite improvement when decoding the MyST test-corpus in case of all the model architectures.
Considering the best results, the children LM provides improvement of 3.87\% relative to adult LM.
However, we find no improvements when testing on OGI Kids corpus.
In context with the perplexity analysis presented in section~\ref{sec:lm_perp}, the reduction in WER is minimal although large improvements were observed in perplexity values on MyST corpus.
This finding can be attributed to two factors: (i) the end-to-end architectures have the ability to implicitly learn language provided enough speech data, and (ii) the acoustic variability in children dominates in our setup.

\section{RESULTS: Children Acoustic Model}\label{sec:results_child_am}
In this section, results are presented on the models fine-tuned on MyST dataset.
All the results also incorporate interpolated language models, i.e., interpolation of language models for LIBRISPEECH and training subset of MyST dataset with interpolation weights tuned on MyST development dataset.
The results are listed in Table~\ref{tab:res_myst}.
Comparing the results with respect to the adult acoustic models (Table~\ref{tab:res_librivox}), we observe a significant performance boost for evaluations made on in-domain MyST test corpus.
The LER of the best performing model improves by a relative 41.63\% and WER by 34.01\%.
Moreover, we also find significant improvements on out-of-domain evaluations made on OGI Kids, an improvement of 11.31\% LER and 9.52\% WER.
We note that improvements on OGI Kids corpus are much lesser than improvements on the in-domain MyST test set.
This observation can be explained with the fact that in-domain testing has matched age demographics of children, whereas with the out-of-domain OGI Kids corpus have a wider, more diverse age demographics.
Another important observation is that even with adaptation on child speech and in-domain evaluations, the performance of children ASR remains much worse (11.1 times worse LER and 6.4 times worse WER) than the adult speech recognition with end-to-end ASR systems.

\subsection{DNN-HMM versus End-to-End Models}
After adaptation to children speech, the DNN-HMM model improves by a relative 56.75\% LER and 59.27\% WER on in-domain MyST test set.
Comparing this to the end-to-end systems (relative improvements of 41.63\% LER and 34.01\% WER), the DNN-HMM system is able to adapt to a greater degree, although in terms of absolute error rates the end-to-end ASR systems outperform the DNN-HMM systems by relative 21.42\% in terms of LER and 17.94\% in terms of WER.
With the OGI Kids corpus the end-to-end ASR systems outperform the DNN-HMM system by 27.01\% (relative) in terms of LER and 24.81\% in terms of WER.

\begin{table*}[t]
\centering
\begin{tabular}{clcclll}
\toprule
& ASR Model & \% Total Error & \% Correct & \% Substitution & \% Deletions & \% Insertions \\
\midrule
\multirow{6}{*}{MyST Test} & TDNN-F DNN-HMM & 47.9 & 63.6 & 68.3 (32.7) & 7.7 (3.7) & 24.2 (11.6) \\
& Transformer + CTC (Greedy) & 25.46 & 78.4 & 56.6 (14.4) & 28.0 (7.1) & 15.3 (3.9) \\
& Transformer + S2S (Greedy) & 29.01 & 77.5 & 51.0 (14.8) & 26.9 (7.8) & 22.4 (6.5) \\
& Transformer + CTC + 4-gram & 25.21 & 77.3 & 46.0 (11.6) & 44.0 (11.1) & 9.9 (2.5) \\
& TDS + S2S + 6-gram-wp & 31.93 & 74.3 & 51.7 (16.5) & 28.8 (9.2) & 18.7 (6.3) \\
& Transformer + CTC + GCNN & 24.26 & 78.5 & 47.4 (11.5) & 41.2 (10.0) & 11.5 (2.8) \\
\arrayrulecolor{black!50}\midrule
\multirow{4}{*}{OGI Kids} & TDNN-F DNN-HMM & 53.55 & 52.5 & 76.0 (40.7) & 12.9 (6.9) & 11.2 (6.0) \\
& Resnet + CTC (Greedy) & 38.00 & 63.5 & 57.9 (22.0) & 38.2 (14.5) & 3.9 (1.5) \\
& Resnet + CTC + 4-gram & 37.32 & 65.3 & 59.8 (22.3) & 33.2 (12.4) & 7.0 (2.6) \\
& TDS + S2S + 6-gram-wp & 74.62 & 26.9 & 21.0 (15.7) & 76.9 (57.4) & 2.1 (1.6) \\
\arrayrulecolor{black}\bottomrule
\end{tabular}
\caption{Word Level Error Analysis of Adult ASR Models on Children's speech\\
\footnotesize{Percent Correct refers to the fraction of the words in the reference that are present in the ASR hypothesis.
For the substitutions, deletions and insertions, the numbers indicate the proportion respective to the total error and the numbers inside the parenthesis are the absolute values.}}\label{tab:analysis:word}
\end{table*}

\begin{table*}[t]
\centering
\begin{tabular}{clcclll}
\toprule
& ASR Model & \% Total Error & \% Correct & \% Substitution & \% Deletions & \% Insertions \\
\midrule
\multirow{6}{*}{MyST Test} & TDNN-F DNN-HMM & 27.0 & 84.0 & 36.7 (9.9) & 22.6 (6.1) & 40.7 (11.0) \\
& Transformer + CTC (Greedy) & 15.71 & 87.8 & 21.6 (3.4) & 56.0 (8.8) & 22.3 (3.5) \\
& Transformer + S2S (Greedy) & 18.78 & 86.5 & 21.3 (4.0) & 50.6 (9.5) & 27.7 (5.2) \\
& Transformer + CTC + 4-gram & 17.6 & 84.7 & 14.8 (2.6) & 71.6 (12.6) & 13.1 (2.3) \\
& TDS + S2S + 6-gram-wp & 21.05 & 84.4 & 21.9 (4.6) & 52.3 (11.0) & 26.1 (5.5) \\
& Transformer + CTC + GCNN & 16.79 & 85.7 & 16.1 (2.7) & 69.7 (11.7) & 14.9 (2.5) \\
\arrayrulecolor{black!50}\midrule
\multirow{4}{*}{OGI Kids} & TDNN-F DNN-HMM & 36.04 & 76.4 & 41.3 (14.9) & 24.1 (8.7) & 34.4 (12.4) \\ 
& Resnet + CTC (Greedy) & 25.75 & 76.8 & 26.4 (6.8) & 64.1 (16.5) & 10.1 (2.6) \\
& Resnet + CTC + 4-gram & 25.02 & 77.6 & 27.2 (6.8) & 62.4 (15.6) & 10.8 (2.7) \\
& TDS + S2S + 6-gram-wp & 71.94 & 29.5 & 7.9 (5.7) & 90.2 (64.9) & 2.1 (1.5) \\
\arrayrulecolor{black}\bottomrule
\end{tabular}
\caption{Character Level Error Analysis of Adult ASR Models on Children speech\\
\footnotesize{Percent Correct refers to the fraction of the words in the reference that are present in the ASR hypothesis.
For the substitutions, deletions and insertions, the numbers indicate the proportion respective to the total error and the numbers inside the parenthesis are the absolute values.}}\label{tab:analysis:char}
\end{table*}

\subsection{Greedy Decoding versus Beamsearch Decoding}
Interestingly, the best performance on the in-domain MyST test data set is obtained with greedy decoding.
This suggests that the inherent language model estimated by the end-to-end systems trained on more than 58,000 hours of adult speech contain sufficient information for processing the children speech in our experiments.
This means that improvements obtained on in-domain dataset after adaptation is all attributed to the acoustics.
This finding also hints that the dominating factor of mismatch between adults and children maybe acoustics.
Overall, the improvements with Greedy decoding is 4.36\% relative WER (10.01\% LER) over beamsearch decoding.

However, for the out-of-domain evaluation on OGI Kids, the best result is obtained with beamsearch decoding.
This could suggest that with heightened acoustic (domain) mismatches, the language model's role becomes more prominent.
The improvements obtained with beamsearch decoding is 2.66\% WER relative to greedy decoding, however the greedy decoding gives a better LER (relative reduction of 6.61\%).

\subsection{End-to-End Architectures}
The Transformer networks give significantly better error rates on in-domain evaluations (MyST test corpus) over the ResNets and TDS Networks.
However, for out-of-domain evaluations on OGI Kids corpus, both the ResNets as well as the TDS networks outperform Transformer networks.
The Transformer networks undergo notable degradation when tested on OGI Kids hinting at generalization issue.
The above observations agree with those made with adult acoustic models under section~\ref{ssec:e2e_arch}.

\begin{table*}[t]
\centering
\begin{tabular}{clcclll}
\toprule
& ASR Model & \% Total Error & \% Correct & \% Substitution & \% Deletions & \% Insertions \\
\midrule
\multirow{5}{*}{MyST Test} & TDNN-F DNN-HMM & 19.51 & 83.9 & 52.3 (10.2) & 30.2 (5.9) & 17.4 (3.4) \\
& Transformer + CTC (Greedy) & 16.01 & 86.5 & 53.1 (8.5) & 31.2 (5.0) & 15.6 (2.5) \\
& Transformer + S2S (Greedy) & 16.69 & 86.5 & 38.3 (6.4) & 41.9 (7.0) & 19.2 (3.2) \\
& Transformer + CTC + 4-gram & 16.74 & 86.5 & 48.4 (8.1) & 32.3 (5.4) & 19.1 (3.2) \\
& Transformer + S2S + 6-gram-wp & 16.81 & 86.7 & 38.7 (6.5) & 40.5 (6.8) & 20.8 (3.5) \\
\arrayrulecolor{black!50}\midrule
\multirow{4}{*}{OGI Kids} & TDNN-F DNN-HMM & 44.74 & 57.2 & 53.6 (24.0) & 42.0 (18.8) & 4.5 (2.0) \\
& TDS + CTC (Greedy) & 34.56 & 67.1 & 59.0 (20.4) & 36.2 (12.5) & 4.6 (1.6) \\
& TDS + S2S (Greedy) & 69.36 & 32.1 & 20.8 (14.4) & 77.1 (53.5) & 2.2 (1.5) \\
& TDS + CTC + 4-gram & 33.64 & 67.6 & 50.2 (16.9) & 46.1 (15.5) & 3.9 (1.3) \\
& TDS + S2S + 4-gram & 64.75 & 37.7 & 26.9 (17.4) & 69.3 (44.9) & 3.9 (2.5) \\
\arrayrulecolor{black}\bottomrule
\end{tabular}
\caption{Word Level Error Analysis of Adapted ASR Models on Children's speech\\
\footnotesize{Percent Correct refers to the fraction of the words in the reference that are present in the ASR hypothesis.
For the substitutions, deletions and insertions, the numbers indicate the proportion respective to the total error and the numbers inside the parenthesis are the absolute values.}}\label{tab:analysis_adapted:word}
\end{table*}

\begin{table*}[t]
\centering
\begin{tabular}{clcclll}
\toprule
& ASR Model & \% Total Error & \% Correct & \% Substitution & \% Deletions & \% Insertions \\
\midrule
\multirow{5}{*}{MyST Test} & TDNN-F DNN-HMM & 13.10 & 89.9 & 27.5 (3.6) & 49.6 (6.5) & 22.9 (3.0) \\
& Transformer + CTC (Greedy) & 9.17 & 93.5 & 15.3 (1.4) & 55.6 (5.1) & 29.4 (2.7) \\
& Transformer + S2S (Greedy) & 11.67 & 91.5 & 13.7 (1.6) & 59.1 (6.9) & 27.4 (3.2) \\
& Transformer + CTC + 4-gram & 10.19 & 92.3 & 14.7 (1.5) & 61.8 (6.3) & 24.5 (2.5) \\
& Transformer + S2S + 6-gram-wp & 11.78 & 91.6 & 13.6 (1.6) & 56.9 (6.7) & 28.9 (3.4) \\
\arrayrulecolor{black!50}\midrule
\multirow{4}{*}{OGI Kids} & TDNN-F DNN-HMM & 30.40 & 73.1 & 27.0 (8.2) & 61.5 (18.7) & 11.5 (3.5) \\
& TDS + CTC (Greedy) & 22.19 & 80.7 & 25.2 (5.6) & 61.7 (13.7) & 11.1 (2.9) \\
& TDS + S2S (Greedy) & 64.75 & 36.9 & 8.3 (5.4) & 89.1 (57.7) & 2.6 (1.7) \\
& TDS + CTC + 4-gram & 23.76 & 78.2 & 19.4 (4.6) & 72.4 (17.2) & 8.4 (2.0) \\
& TDS + S2S + 4-gram & 59.27 & 43.8 & 11.8 (7.0) & 83.0 (49.2) & 5.2 (3.1) \\
\arrayrulecolor{black}\bottomrule
\end{tabular}
\caption{Char Level Error Analysis of Adapted ASR Models on Children speech\\
\footnotesize{Percent Correct refers to the fraction of the words in the reference that are present in the ASR hypothesis.
For the substitutions, deletions and insertions, the numbers indicate the proportion respective to the total error and the numbers inside the parenthesis are the absolute values.}}\label{tab:analysis_adapted:char}
\end{table*}

\subsection{CTC versus Sequence-to-Sequence Training}
As observed with the adult acoustic models (under section~\ref{ssec:ctc_vs_s2s}), the sequence-to-sequence models are always outperformed by the CTC loss training both with in-domain evaluation on MyST corpus as well as with the OGI Kids corpus.
The performance difference between the CTC training and the sequence-to-sequence increase on out-of-domain OGI Kids corpus.
However, the difference remains much lower with in-domain testing on MyST corpus.
We believe the matched age demographics with in-domain MyST testing helps the sequence-to-sequence models.
Overall, we find the sequence-to-sequence training to be less generalizable for children speech recognition.

\subsection{Language Models}
In Table~\ref{tab:res_myst}, we note that most of the best results are obtained with greedy decoding, i.e., without a LM.
This is in contrast to the improvements that were noted with LM for adult AM seen in Table~\ref{tab:res_librispeech}.
Regardless of the large improvements in perplexity on MyST corpus with inclusion of children LM, see section~\ref{sec:lm_perp}, we find no improvements with beamsearch decoding.
This suggests that the end-to-end models are capable of modeling language given enough training data.
It also indicates that acoustic mismatch is the dominating factor for children speech and addressing it is responsible for most of the gains with children speech recognition.
This observation is in agreement with the study in \cite{shivakumar2020transfer}, where transfer learning of layers close to acoustic features accounted for the maximum improvements suggesting acoustic variability is the dominating factor for degradation in children speech recognition.

\section{Error Analysis}\label{sec:error_analysis}
In this section, we conduct various analyses to get further insights into errors made by the aforementioned ASR systems.

\subsection{Error Rate Analysis}
We conduct a breakdown of the error rates in terms of substitution, deletions and insertions to assess the strengths and weakness of DNN-HMM and end-to-end systems, as well as different architectures and loss functions.
Table~\ref{tab:analysis:word} shows the breakdown of the WER of various acoustic models trained on adults speech.
The choice of the models are such that we cover different aspects such as greedy versus beamsearch, CTC versus sequence-to-sequence, DNN-HMM versus end-to-end systems.
From Table~\ref{tab:analysis:word}, we find that substitutions and insertions are more suppressed with the end-to-end systems compared to the DNN-HMM system while the deletions are inflated.
We find this trend to be consistent across both MyST corpus as well as the OGI Kids corpus and over various configurations including greedy, beamsearch decoding, CTC and Sequence-to-sequence training and various language models.
We observe that in the case of breakdown of sequence-to-sequence models, there is a big spike for deletions.
All the above observations are prevalent even at the level of characters, with error rate analysis presented in Table~\ref{tab:analysis:char}.

Error rate analysis for the acoustic models trained on MyST kids corpus are presented in Table~\ref{tab:analysis_adapted:word} and Table~\ref{tab:analysis_adapted:char}.
After adaptation with children speech, we observe the proportion of deletions of DNN-HMM system increases and the insertions decrease and becomes more comparable with that of end-to-end systems (see Table~\ref{tab:analysis_adapted:word}).
The deletions of the end-to-end system continue to be more than the DNN-HMM systems, whereas the substitutions remain relatively low.
The above observations is consistent across both MyST corpus and the OGI Kids corpus and also with character level error analysis in Table~\ref{tab:analysis_adapted:char}.

\begin{figure}[t]
\includegraphics[width=\columnwidth]{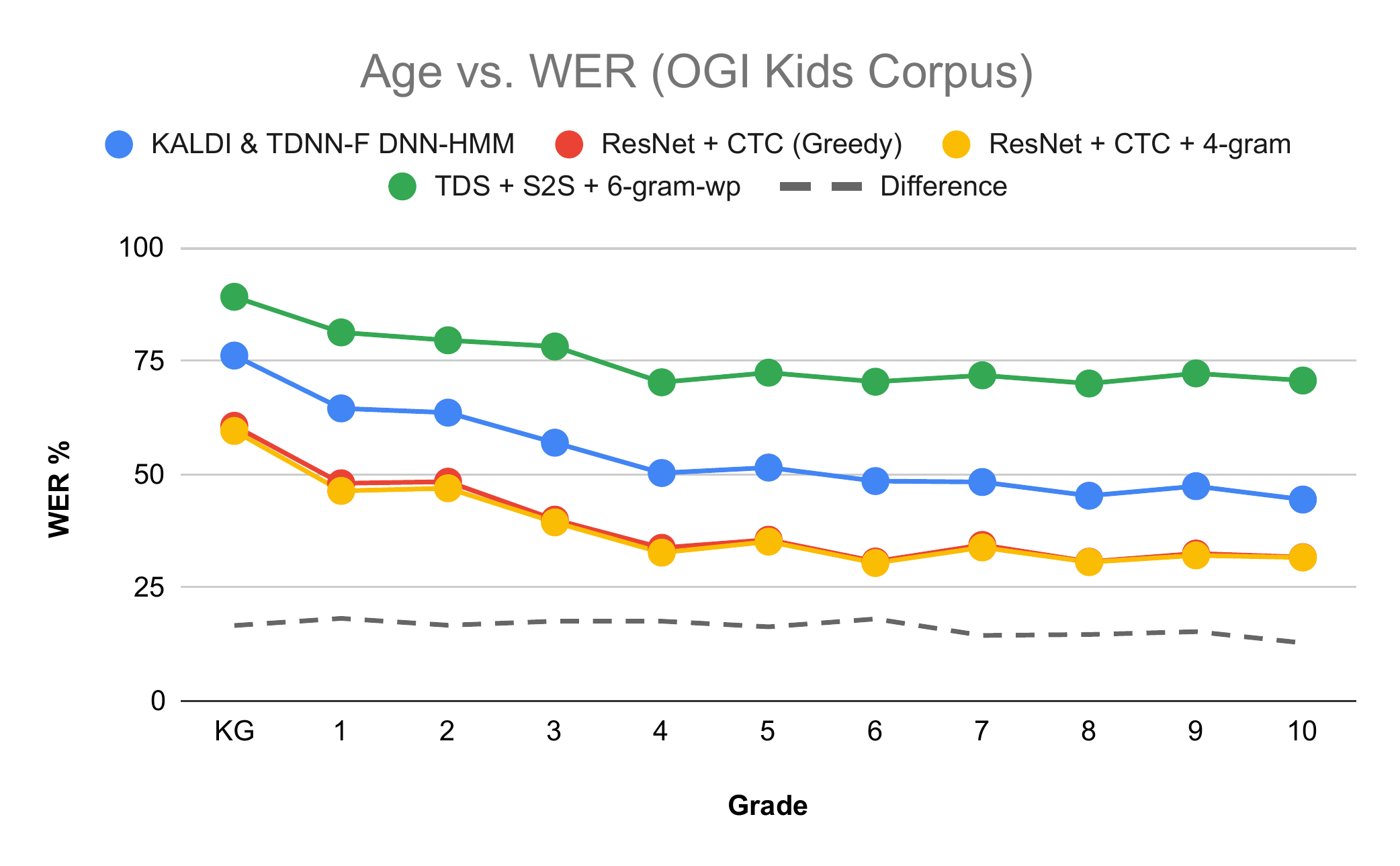}
\caption{Age versus WER for Adult AM trained on LIBRISPEECH + LIBRIVOX}\label{fig:age_vs_wer_adult}
\end{figure}

\begin{figure}[t]
\includegraphics[width=\columnwidth]{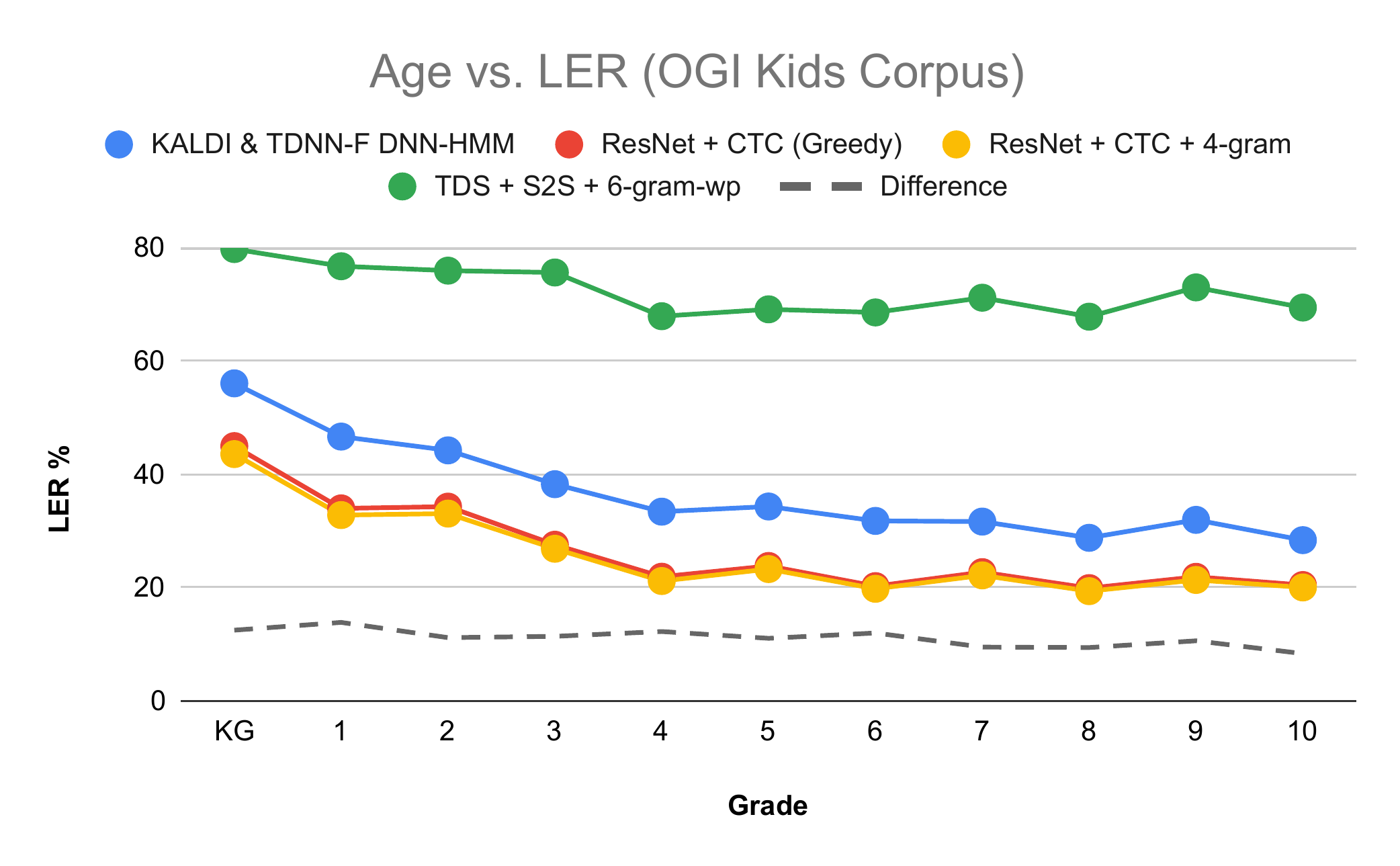}
\caption{Age versus LER for Adult AM trained on LIBRISPEECH + LIBRIVOX}\label{fig:age_vs_ler_adult}
\end{figure}

\subsection{Effect of Age}
In this section, we assess the error rates with respect to children's age.
All the age related evaluations are performed on the OGI Kids Corpus since it has diverse age distribution among children.
\subsubsection{Adult Acoustic Models}
Figure~\ref{fig:age_vs_wer_adult} plots the WER obtained on adult acoustic model trained on combination of LIBRISPEECH and LIBRIVOX, corresponding to Table~\ref{tab:res_librivox} for OGI Kids Corpus across school grades.
Firstly, we observe that the WER is worst for kindergarten children and gets progressively better with increase in children's age.
The decrease in WER is steep until 4th grade and relatively flattens out.
The above trends are consistent over all the model configurations including DNN-HMM, greedy and beamsearch decoding, CTC and sequence-to-sequence networks.
In sum, the age associated challenges with children speech recognition are prevalent even in the end-to-end systems and their trends are similar to previous works involving GMM-HMM systems \cite{shivakumar2014improving} and DNN-HMM systems \cite{shivakumar2020transfer}.

Comparing the DNN-HMM based model with the best-performing ResNets based end-to-end system, nearly constant improvements are obtained with the end-to-end system over all age categories.
The difference between the error rates between the TDNN-F HMM and the end-to-end systems is minimal for eldest children (10th grade).
We do not observe any striking differences between different architectures and loss functions of the end-to-end systems.
Moreover, plots of letter error rate in Figure~\ref{fig:age_vs_ler_adult} also agree with the earlier observations.

\begin{figure}[t]
\includegraphics[width=\columnwidth]{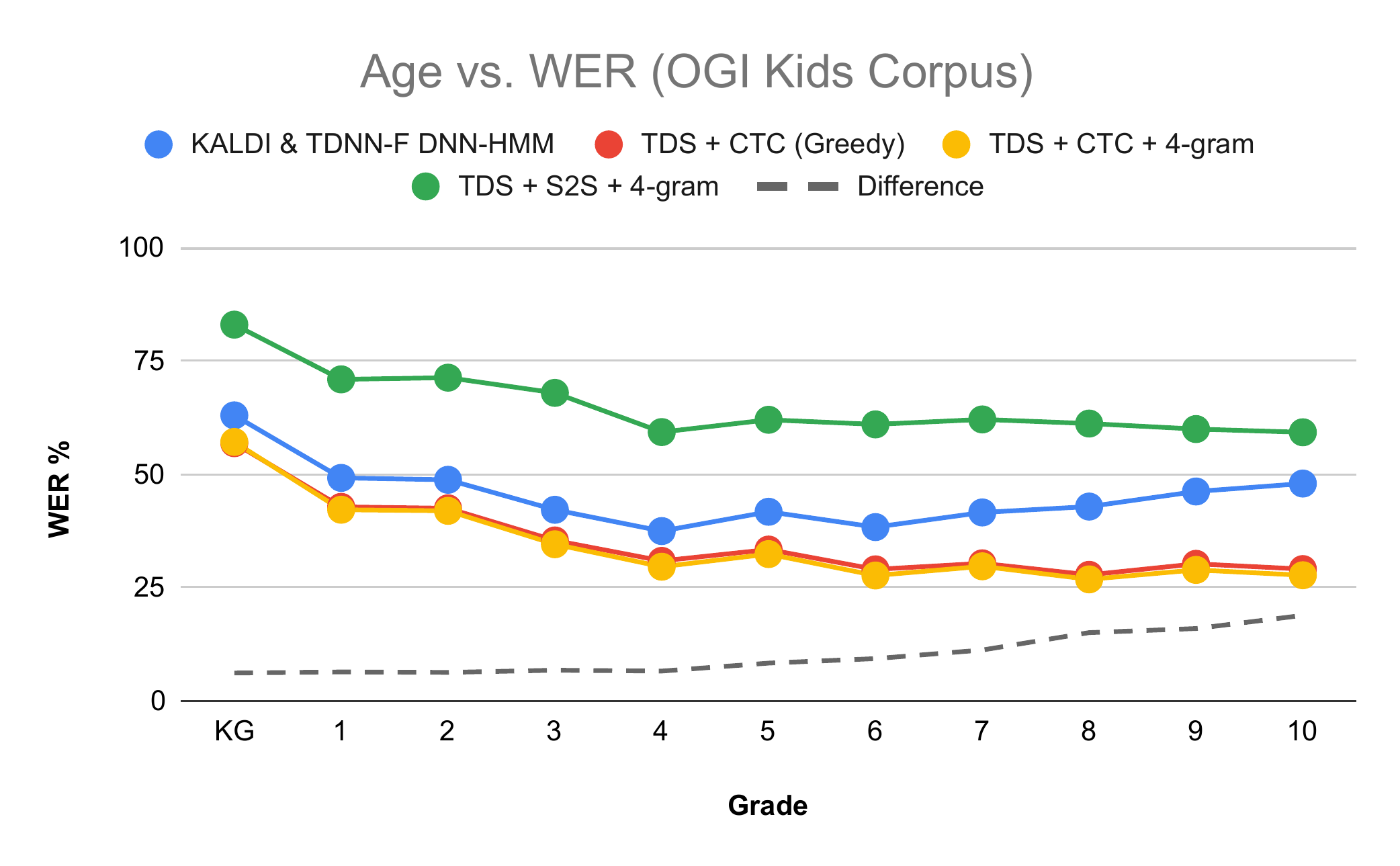}
\caption{Age versus WER for AM fine-tuned on MyST}\label{fig:age_vs_wer_kids}
\end{figure}

\begin{figure}[t]
\includegraphics[width=\columnwidth]{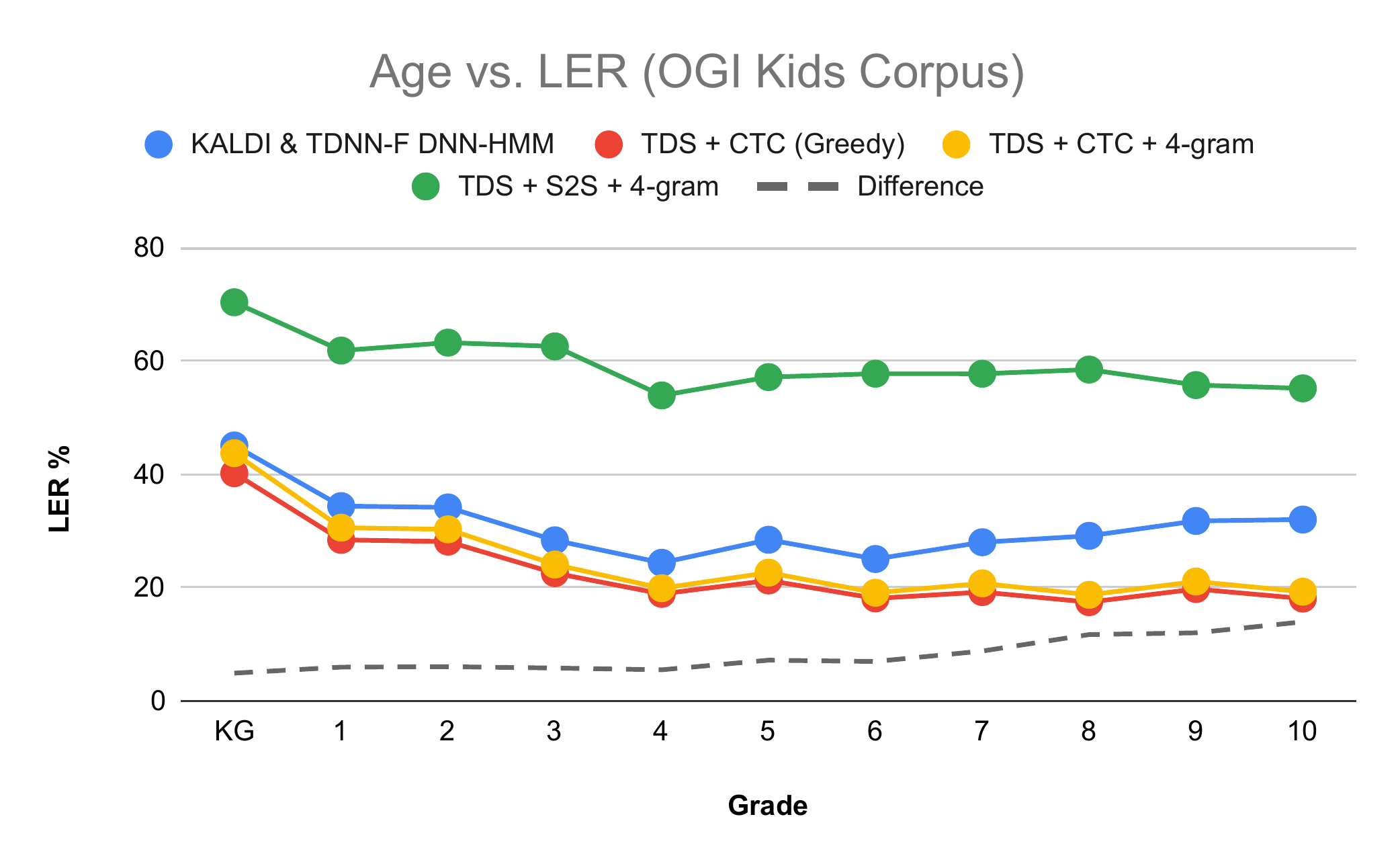}
\caption{Age versus LER for AM fine-tuned on MyST}\label{fig:age_vs_ler_kids}
\end{figure}

\subsubsection{Children Acoustic Models}
Figure~\ref{fig:age_vs_wer_kids} plots the WER obtained on acoustic models adapted on MyST corpus.
Note, the acoustic models were adapted with data corresponding to children studying in grades 3 to 5.
Similar to observation with adult acoustic models, we find the WER is worst for kindergarten children.
For end-to-end architectures, the WER decreases steeply until grade 4 and flattens out just as in the case of the adult acoustic model.
Interestingly, we find that the trends observed with end-to-end architectures are nearly identical as was observed in the unadapted baseline adult models despite training on children of grades 3 - 5.
With the end-to-end architecture there is near constant improvements in absolute WER throughout all the ages in spite of adapting on data from only a subset of age categories (grades 3 -5).
However, with the TDNN-F DNN-HMM model we see that the WER decreases and reaches minimum for grade 4 and we observe an increase in WER for children of grade 7 and above.
This suggests that the DNN-HMM models are more sensitive to the children's age i.e., adaptation data age range.

\begin{figure}[t]
\includegraphics[width=\columnwidth]{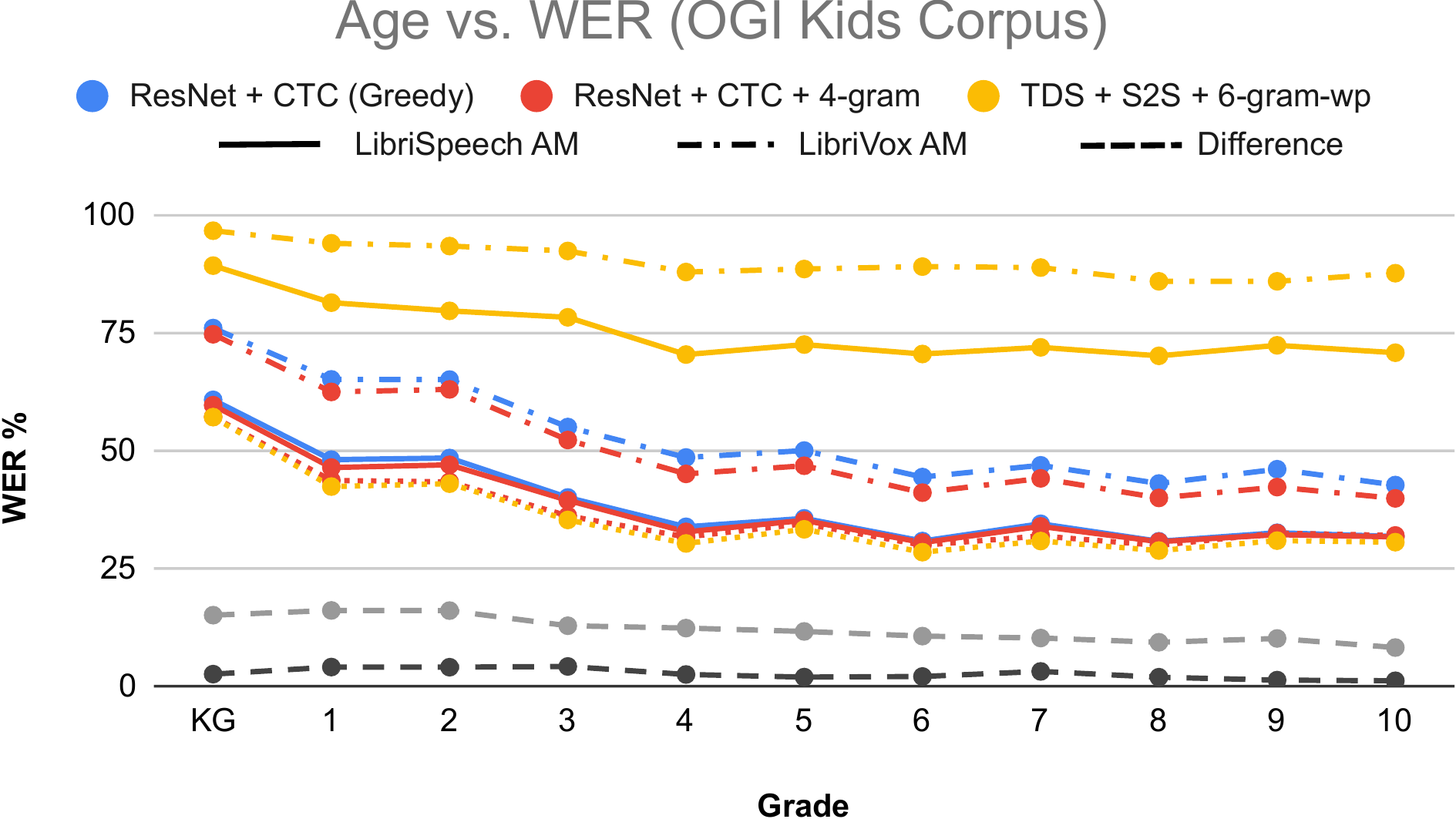}
\caption{Age versus WER for AM trained on different amounts of data}\label{fig:age_vs_wer_data}
\end{figure}

Comparing the DNN-HMM acoustic models with the end-to-end architectures, we note the WER with TDNN-F HMM for kindergarten children improves by 17.31\% relative to adult acoustic models and the WER with the best performing end-to-end acoustic model (TDS + CTC + 4-gram LM) for kindergarten children improves only by a relative 4.08\%.
The differences in WER between the DNN-HMM models and end-to-end system interestingly increase with children's age.
No major differences were observed between different architectures of end-to-end system (CTC vs. sequence-to-sequence and Greedy vs. beamsearch decoding).

Figure~\ref{fig:age_vs_ler_kids} plots the LER obtained with acoustic models adapted on MyST corpus.
Few notable differences can be spotted relative to the earlier observed trends with WERs.
First, we observe that the greedy decoding yields better LER in comparison with its beamsearch counterpart, and these improvements are throughout all ages.
Second, we note that with the beamsearch decoding, the LER is relatively worse for younger children, i.e., the greedy decoding is better in terms of LER for children of kindergarten to grade 3.
The above two observations, suggest that the language model hampers the LER in adapted models while providing no improvements in terms of WER.
Finally, we observe that the difference in LER between the DNN-HMM and the best performing end-to-end model (TDS + CTC) with beamsearch decoding is minimal.

\subsection{Effect of Data}
\begin{figure}[t]
\includegraphics[width=\columnwidth]{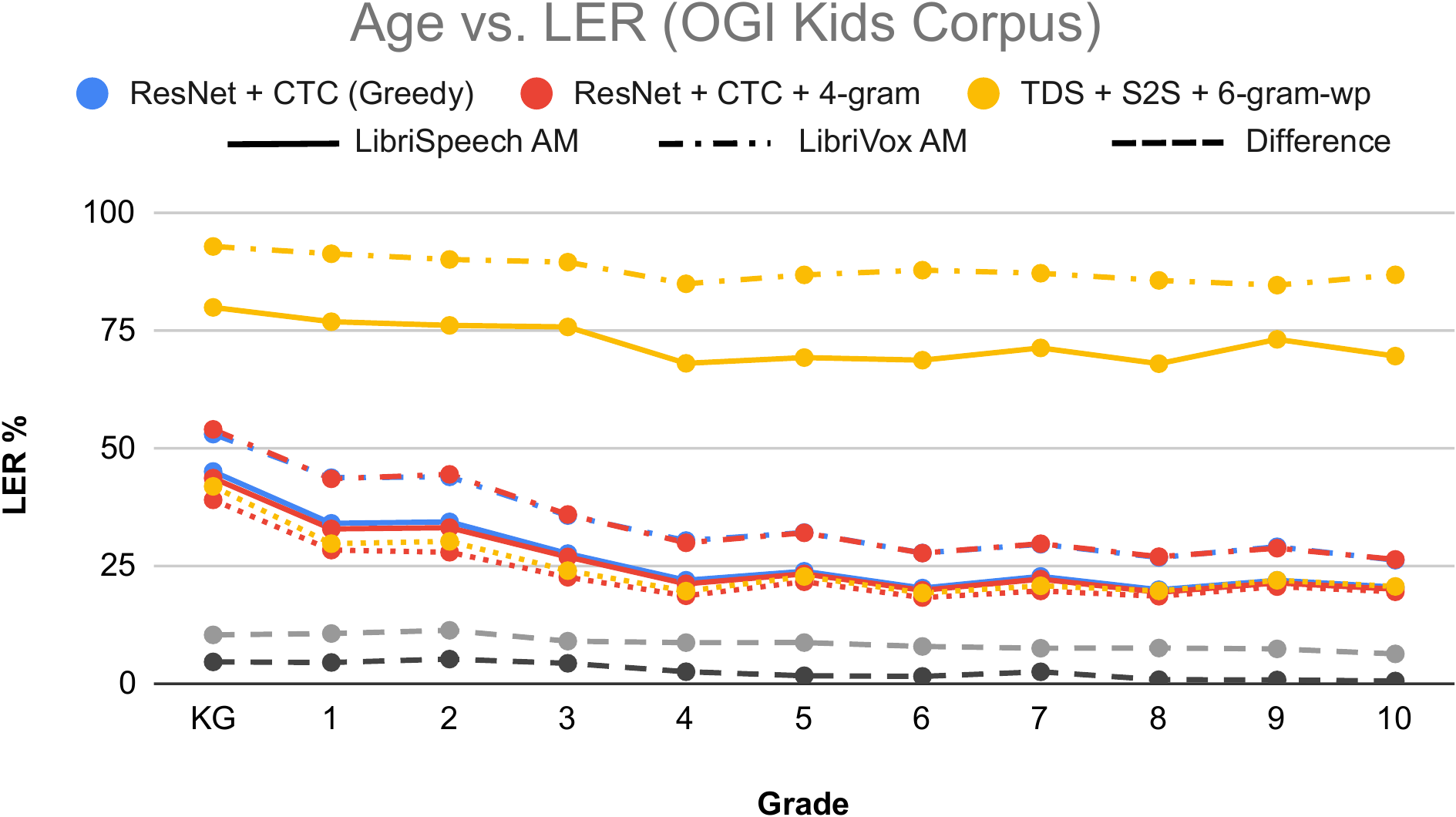}
\caption{Age versus LER for AM trained on different amounts of data}\label{fig:age_vs_ler_data}
\end{figure}

In this section, we analyze the effect of training data on the performance over different age categories.
We consider acoustic models trained on: (i) LIBRISPEECH, (ii) LIBRISPEECH + LIBRIVOX, and (iii) LIBRISPEECH + LIBRIVOX adapted on MyST corpus.
Figure~\ref{fig:age_vs_wer_data} and Figure~\ref{fig:age_vs_ler_data} illustrates the plots of WER and LER over different age categories respectively.
First, we observe that addition of 52,700 hours of LIBRIVOX data helps lower WER over all the age categories by a considerable margin.
An important observation here is that with the addition of large amounts of adult speech data for training, relatively larger improvements are observed for younger children (kindergarten to 3 grade) compared to elder children.
Further adaptation with children speech data mainly helps the speech recognition for younger children and does not seem to provide significant improvements for older children. 
Noting that these results are on out-of-domain OGI Kids corpus, we find relative improvements in WER of 20.15\% with an increase of 54 times of adult training data.
Whereas, the relative improvements of 4.24\% is obtained with just 0.37\% of children speech data for kindergarten children.

\begin{figure}[t]
\includegraphics[width=\columnwidth]{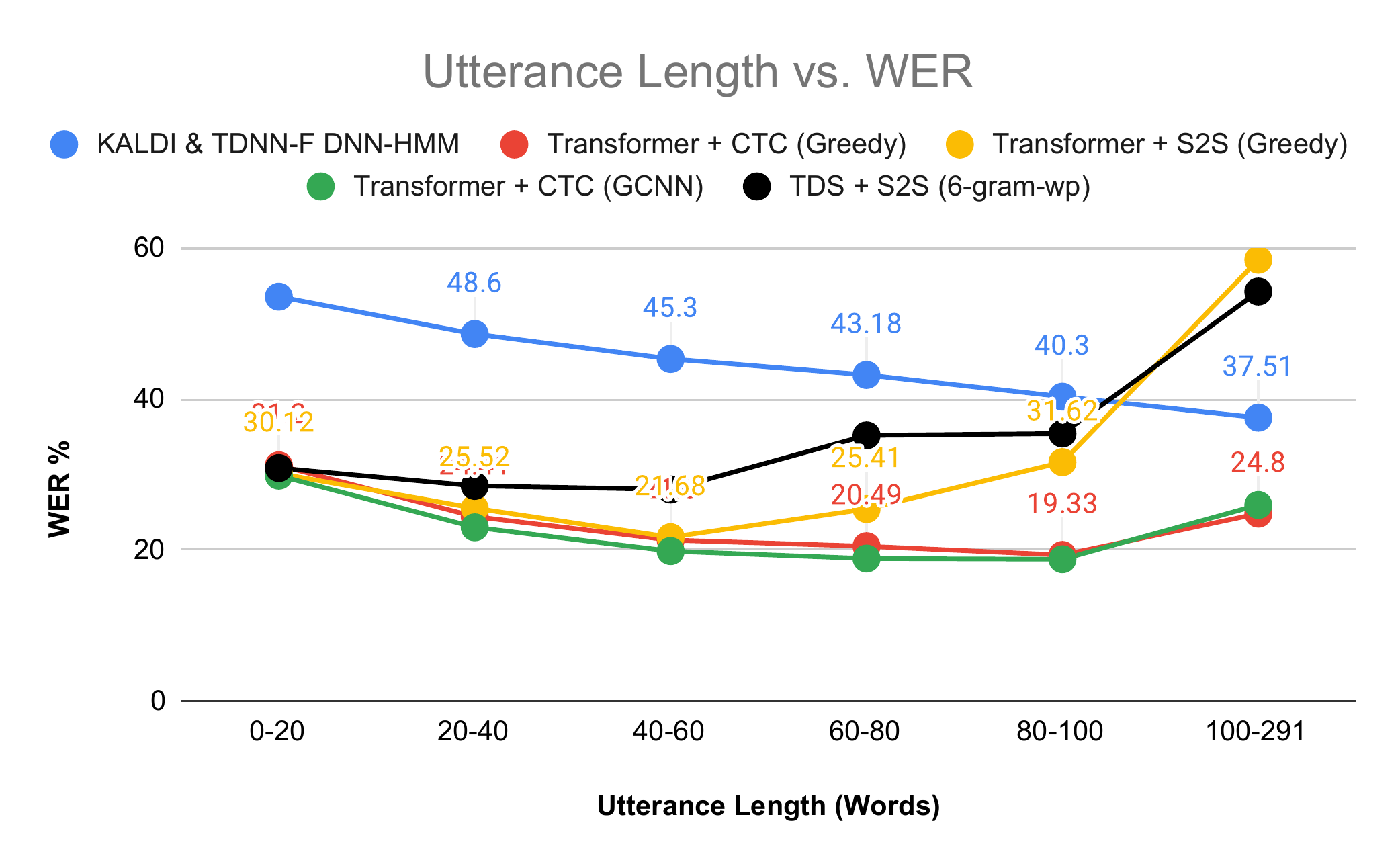}
\caption{Utterance Length versus WER for Adult AM trained on LIBRISPEECH + LIBRIVOX}\label{fig:myst_length_vs_wer_adult}
\end{figure}

\begin{figure}[t]
\includegraphics[width=\columnwidth]{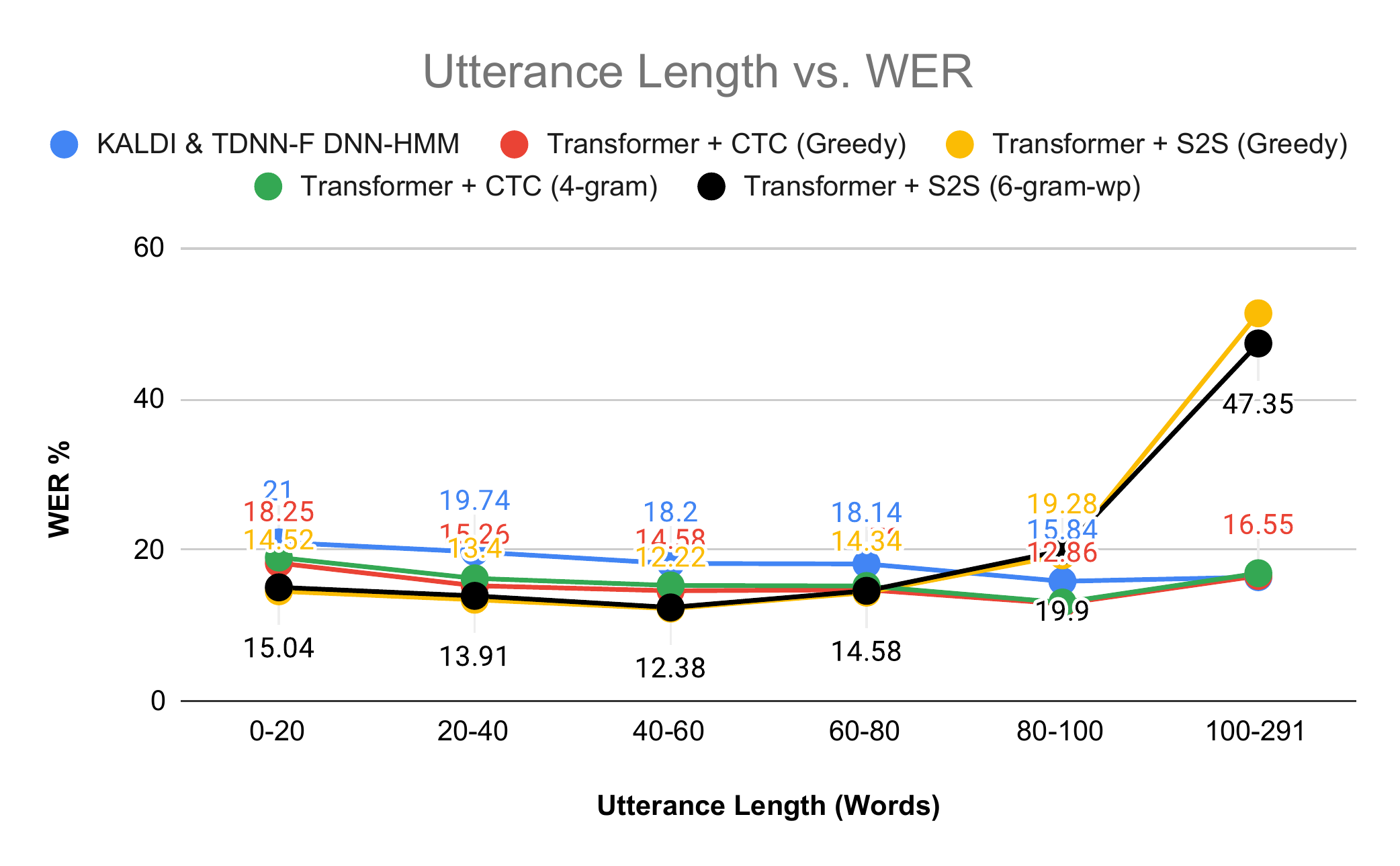}
\caption{Utterance Length versus WER for AM fine-tuned on MyST}\label{fig:myst_length_vs_wer_kids}
\end{figure}

\begin{figure}[!b]
\includegraphics[width=\columnwidth]{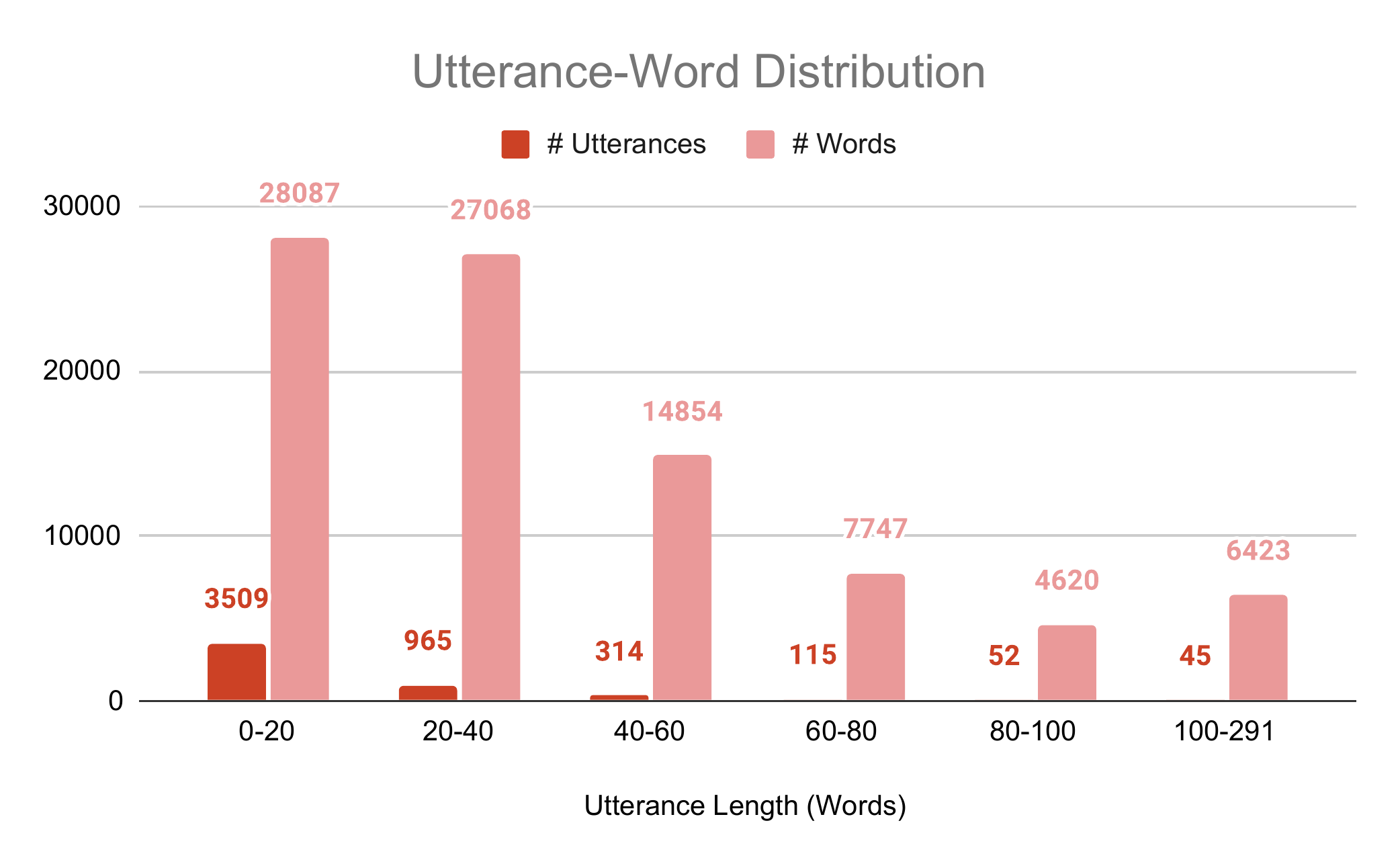}
\caption{Utterance-Word Length Distribution MyST Corpus}\label{fig:myst_length_dist}
\end{figure}

\subsection{Effect of Utterance Length}
Here we analyze the effect of utterance length on the performance of the various acoustic models.
The utterance length distribution of the MyST test subset is shown in Figure~\ref{fig:myst_length_dist}.
Figure~\ref{fig:myst_length_vs_wer_adult} illustrates the plot of WER on MyST test corpus with adult acoustic model for utterances of varying lengths.
We observe the performance of TDNN-F based DNN-HMM system improves as the utterance length increases.
However, with the end-to-end systems we see degradation for longer utterance lengths.
The performance of the CTC based system improves initially with increase in utterance lengths and then degrades for utterance lengths of over 100 words.
The sequence-to-sequence acoustic models undergo a more drastic degradation for utterance lengths over 60 words.
While the GCNN LM provides slight improvements for utterance lengths under 80 words, they provide no advantage for longer utterances.

\begin{table*}[t]
\centering
\begin{tabular}{lll}
\toprule
Language Model & Perplexity (MyST-test) & Perplexity (OGI Kids) \\
\midrule
4-gram \footnotesize{LIBRISPEECH} & 574.86 & 281.51 \\
6-gram-wp \footnotesize{LIBRISPEECH} & 216.87 & 126.05 \\
GCNN \footnotesize{LIBRISPEECH} & 174.89 & 82.88 \\
GCNN-wp \footnotesize{LIBRISPEECH} & 105.29 & 76.50 \\
3-gram \footnotesize{MyST} & 102.16 & 300.20 \\
6-gram-wp \footnotesize{MyST} & 85.61 & 210.43 \\
4-gram \footnotesize{LIBRISPEECH + MyST} & 92.52 & 170.58 \\
6-gram-wp \footnotesize{LIBRISPEECH + MyST} & 65.25 & 140.68 \\
GCNN \footnotesize{LIBRISPEECH + MyST} & 187.88 & 134.49 \\
\bottomrule
\end{tabular}
\caption{Language Model Perplexities}\label{tab:lm_perp}
\end{table*}

Figure~\ref{fig:myst_length_vs_wer_kids} plots the WER on MyST test corpus with the in-domain adapted acoustic model over varying utterance lengths.
Similar to the observations made with the adult acoustic model, the DNN-HMM system's performance improves with length and does not undergo any degradation for longer utterances.
We note that compared to Figure~\ref{fig:myst_length_vs_wer_adult}, the improvements are relatively low with increase in length.
Next, with the end-to-end architecture with CTC training, we observe slight degradation for utterances of over 100 words in length.
Comparatively, the degradation is of lower magnitude to the one observed with adult acoustic model (see Figure~\ref{fig:myst_length_vs_wer_adult}).
The degradation is drastic for sequence-to-sequence architectures for utterance lengths greater than 80.
Comparatively, the degradation onset is greater, however more acute to the one observed with adult acoustic model (see Figure~\ref{fig:myst_length_vs_wer_adult}).
Interestingly, the sequence-to-sequence architecture performs better than the CTC trained counterpart under utterance lengths of 80 with adapted acoustic models.
Another important observation is that the performance of the DNN-HMM system equals to that of the best performance end-to-end systems for longer utterances (greater than 100 words).

\subsection{Language Model Perplexities}\label{sec:lm_perp}
Table~\ref{tab:lm_perp} presents the perplexity of the various language models on MyST and OGI Kids corpus.
The perplexities provide more context to analyze the results in Table~\ref{tab:res_librivox} and Table~\ref{tab:res_myst}.
Comparing different LMs we find the word-piece 6-gram models provide reduction in perplexity of more than 50\% over the traditional 4-gram word based LM.
The gated convolutional neural network LM yield perplexities 70\% lower than typical 4-gram word LMs.
The word-piece based GCNN LM provides further improvements and gives the lowest perplexities.
Among the two children speech corpora, we observe OGI Kids has higher perplexities compared to the MyST corpus and this also reflects in terms of WER in previous assessments.
We also find original language model trained on public domain books show higher degree of mismatch with children corpus \cite{zeghidour2019fully}.
The addition of children LM noticeably reduces the perplexity for both word-based and word-piece based n-gram models on MyST corpus.
However, the inclusion of children data does not help the perplexities on the OGI Kids corpus, i.e., the perplexities after addition of children LM results in higher perplexities.
In case of GCNN LM, although fine-tuning the LM helps decrease perplexity on development set, we find that they do not translate well to the test set, resulting in slight increase in perplexity values.
Hence, we skip the results of speech recognition with child adapted GCNN LM under Table~\ref{tab:res_librivox} and Table~\ref{tab:res_myst}.

\subsection{Confusion Analysis}
Table~\ref{tab:cp_myst} and Table~\ref{tab:cp_ogi} lists the top-50 confusion pairs among the best performing DNN-HMM, end-to-end models trained on LIBRISPEECH, end-to-end models trained on LIBRIVOX and end-to-end models adapted on children speech corpus, evaluated on MyST and OGI corpus respectively.
Beginning with the TDNN-F model, most of the errors are because of (i) deletion of certain consonants which result in either partial word recognition such as biosphere being recognized as sphere, meat being recognized as me, (ii) substitution/confusion between vowels that result in errors between acoustically similar words such as like \& lake, um \& arm, and (iii) confusion involving fillers including ah, uh, um, uhm etc.
Second observation is that most of the errors are among stop words.
With end-to-end system, trained on LIBRISPEECH, we see suppression of all three kinds of errors but more specifically we notice deletion of consonants to be minimum and the confusion involving the fillers to be less prevalent.
With acoustic models trained on additional speech data, LIBRIVOX, the confusion pairs are similar to the one observed with the LIBRISPEECH model, but there is suppression of errors all along the spectrum.
After adapting the acoustic model on children data, we observe less prevalence of word-confusions among stop words.
Overall, we find the end-to-end models are more efficient in handling filler words and confusions resultant from deletion of consonants or breaking of words such as plant \& plan, about \& bow etc.

\begin{table*}[p]
\centering
\resizebox{\textwidth}{!}{%
\begin{tabular}{rlrlrlrl}
\multicolumn{2}{c}{TDNN-F HMM} & \multicolumn{2}{c}{Transformer + CTC + 4-gram {\tiny(LIBRISPEECH)}} & \multicolumn{2}{c}{Transformer + CTC + GCNN {\tiny(LIBRIVOX)}} & \multicolumn{2}{c}{Transformer + CTC + GCNN {\tiny(MyST)}} \\ 
\cmidrule(lr){1-2} \cmidrule(lr){3-4} \cmidrule(lr){5-6} \cmidrule(lr){7-8}
Frequency & Confusion & Frequency & Confusion & Frequency & Confusion & Frequency & Confusion\\
\cmidrule(lr){1-2} \cmidrule(lr){3-4} \cmidrule(lr){5-6} \cmidrule(lr){7-8}
801 & and $\rightarrow$ in & 234 & and $\rightarrow$ in & 145 & and $\rightarrow$ in & 126 & because $\rightarrow$ cause\\
244 & and $\rightarrow$ an & 107 & um $\rightarrow$ on & 102 & a $\rightarrow$ the & 100 & $<unk>$ $\rightarrow$ decomposers\\
189 & um $\rightarrow$ arm & 103 & a $\rightarrow$ the & 94 & $<unk>$ $\rightarrow$ decompose & 86 & $<unk>$ $\rightarrow$ decomposer\\
148 & the $\rightarrow$ a & 88 & the $\rightarrow$ a & 84 & the $\rightarrow$ a & 80 & $<unk>$ $\rightarrow$ biosphere\\
103 & uh $\rightarrow$ ah & 85 & uh $\rightarrow$ a & 83 & $<unk>$ $\rightarrow$ atmosphere & 80 & a $\rightarrow$ the\\
101 & it's $\rightarrow$ its & 82 & it's $\rightarrow$ its & 76 & $<unk>$ $\rightarrow$ decomposes & 66 & in $\rightarrow$ and\\
86 & it $\rightarrow$ a & 68 & $<unk>$ $\rightarrow$ systems & 75 & $<unk>$ $\rightarrow$ systems & 63 & its $\rightarrow$ it's\\
86 & it's $\rightarrow$ it & 62 & $<unk>$ $\rightarrow$ decomposes & 73 & it's $\rightarrow$ its & 63 & the $\rightarrow$ a\\
78 & the $\rightarrow$ though & 61 & $<unk>$ $\rightarrow$ decomposing & 57 & um $\rightarrow$ a & 61 & and $\rightarrow$ in\\
67 & because $\rightarrow$ cause & 61 & um $\rightarrow$ and & 56 & uh $\rightarrow$ a & 51 & $<unk>$ $\rightarrow$ herbivore\\
56 & like $\rightarrow$ lake & 49 & in $\rightarrow$ and & 52 & um $\rightarrow$ on & 43 & $<unk>$ $\rightarrow$ subsystems\\
55 & um $\rightarrow$ on & 49 & um $\rightarrow$ ah & 50 & $<unk>$ $\rightarrow$ synthesis & 42 & $<unk>$ $\rightarrow$ omnivore\\
54 & a $\rightarrow$ the & 44 & $<unk>$ $\rightarrow$ sphere & 49 & um $\rightarrow$ and & 40 & $<unk>$ $\rightarrow$ the\\
53 & biosphere $\rightarrow$ sphere & 44 & $<unk>$ $\rightarrow$ synthesis & 48 & in $\rightarrow$ and & 38 & $<unk>$ $\rightarrow$ photosynthesis\\
49 & it $\rightarrow$ eh & 44 & um $\rightarrow$ a & 41 & $<unk>$ $\rightarrow$ the & 36 & it's $\rightarrow$ its\\
49 & meat $\rightarrow$ me & 43 & $<unk>$ $\rightarrow$ atmosphere & 41 & it's $\rightarrow$ is & 35 & $<unk>$ $\rightarrow$ o\\
49 & the $\rightarrow$ this & 40 & yeast $\rightarrow$ east & 40 & esophagus $\rightarrow$ oesophagus & 35 & uh $\rightarrow$ u\\
46 & eat $\rightarrow$ e & 39 & two $\rightarrow$ too & 35 & predators $\rightarrow$ creditors & 31 & $<unk>$ $\rightarrow$ omnivores\\
43 & into $\rightarrow$ to & 37 & that $\rightarrow$ the & 34 & nutrients $\rightarrow$ nuts & 26 & $<unk>$ $\rightarrow$ geosphere\\
40 & and $\rightarrow$ a & 36 & uh $\rightarrow$ ah & 34 & they're $\rightarrow$ are & 26 & it $\rightarrow$ it's\\
40 & the $\rightarrow$ thee & 34 & it's $\rightarrow$ is & 33 & um $\rightarrow$ ah & 25 & $<unk>$ $\rightarrow$ c\\
39 & eats $\rightarrow$ eat & 34 & that's $\rightarrow$ that & 32 & cause $\rightarrow$ because & 25 & $<unk>$ $\rightarrow$ hydrosphere\\
37 & plants $\rightarrow$ plant & 33 & $<unk>$ $\rightarrow$ system & 32 & that $\rightarrow$ the & 25 & that $\rightarrow$ the\\
36 & it $\rightarrow$ i & 33 & $<unk>$ $\rightarrow$ the & 32 & two $\rightarrow$ too & 22 & $<unk>$ $\rightarrow$ learned\\
36 & that $\rightarrow$ the & 31 & predators $\rightarrow$ creditors & 31 & $<unk>$ $\rightarrow$ sphere & 21 & bloodstream $\rightarrow$ stream\\
34 & photosynthesis $\rightarrow$ synthesis & 30 & esophagus $\rightarrow$ oesophagus & 30 & $<unk>$ $\rightarrow$ system & 20 & $<unk>$ $\rightarrow$ ecosystem\\
34 & subsystems $\rightarrow$ systems & 30 & the $\rightarrow$ that & 30 & it's $\rightarrow$ it & 20 & it's $\rightarrow$ it\\
33 & the $\rightarrow$ that & 30 & they're $\rightarrow$ are & 29 & palmate $\rightarrow$ palm & 20 & uh $\rightarrow$ a\\
32 & plant $\rightarrow$ plan & 29 & cause $\rightarrow$ because & 28 & $<unk>$ $\rightarrow$ o & 19 & are $\rightarrow$ they're\\
32 & um $\rightarrow$ ah & 28 & $<unk>$ $\rightarrow$ decompose & 26 & there's $\rightarrow$ is & 19 & it's $\rightarrow$ is\\
31 & the $\rightarrow$ their & 28 & eats $\rightarrow$ eat & 24 & eats $\rightarrow$ eat & 18 & cord $\rightarrow$ chord\\
31 & the $\rightarrow$ they & 28 & it's $\rightarrow$ it & 23 & they're $\rightarrow$ their & 18 & the $\rightarrow$ they\\
31 & they're $\rightarrow$ there & 28 & meat $\rightarrow$ me & 20 & chlorophyll $\rightarrow$ chloroform & 17 & $<unk>$ $\rightarrow$ ecosystems\\
30 & they're $\rightarrow$ their & 27 & $<unk>$ $\rightarrow$ decomposed & 20 & it $\rightarrow$ i & 17 & is $\rightarrow$ there's\\
30 & um $\rightarrow$ hum & 27 & it $\rightarrow$ that & 20 & nutrients $\rightarrow$ nut & 15 & $<unk>$ $\rightarrow$ and\\
29 & is $\rightarrow$ as & 24 & systems $\rightarrow$ system & 20 & their $\rightarrow$ the & 15 & notice $\rightarrow$ noticed\\
29 & plants $\rightarrow$ plans & 24 & there's $\rightarrow$ is & 19 & it $\rightarrow$ and & 15 & plant $\rightarrow$ plants\\
29 & that's $\rightarrow$ that & 24 & they're $\rightarrow$ their & 19 & meat $\rightarrow$ me & 15 & the $\rightarrow$ that\\
28 & about $\rightarrow$ bow & 24 & yeah $\rightarrow$ yes & 19 & yeast $\rightarrow$ east & 15 & this $\rightarrow$ the\\
28 & are $\rightarrow$ or & 23 & um $\rightarrow$ i'm & 18 & $<unk>$ $\rightarrow$ a & 14 & they're $\rightarrow$ they\\
28 & cells $\rightarrow$ selves & 22 & it $\rightarrow$ i & 18 & because $\rightarrow$ cause & 13 & $<unk>$ $\rightarrow$ bronchi\\
28 & it's $\rightarrow$ is & 20 & $<unk>$ $\rightarrow$ war & 18 & into $\rightarrow$ to & 13 & that $\rightarrow$ it\\
28 & to $\rightarrow$ too & 20 & is $\rightarrow$ as & 18 & it $\rightarrow$ that & 13 & their $\rightarrow$ the\\
27 & and $\rightarrow$ end & 20 & they're $\rightarrow$ they & 18 & yeah $\rightarrow$ yes & 12 & $<unk>$ $\rightarrow$ is\\
27 & decomposers $\rightarrow$ composers & 19 & nutrients $\rightarrow$ nuts & 17 & $<unk>$ $\rightarrow$ and & 12 & $<unk>$ $\rightarrow$ subsystem\\
27 & it $\rightarrow$ at & 19 & um $\rightarrow$ i & 17 & $<unk>$ $\rightarrow$ bronco & 12 & am $\rightarrow$ i'm\\
27 & maybe $\rightarrow$ be & 18 & is $\rightarrow$ there's & 17 & $<unk>$ $\rightarrow$ decomposing & 12 & eats $\rightarrow$ eat\\
27 & um $\rightarrow$ om & 17 & $<unk>$ $\rightarrow$ a & 17 & plants $\rightarrow$ plant & 12 & or $\rightarrow$ are\\
26 & the $\rightarrow$ de & 17 & are $\rightarrow$ our & 17 & this $\rightarrow$ the & 12 & snake $\rightarrow$ rattlesnake\\
26 & they're $\rightarrow$ are & 17 & it $\rightarrow$ and & 16 & that's $\rightarrow$ that & 12 & to $\rightarrow$ into\\
\bottomrule
\end{tabular}}
\caption{Top 50 Confusion Pairs: On MyST Corpus for top performing DNN-HMM system and end-to-end ASR systems}\label{tab:cp_myst}
\end{table*}

\begin{table*}[p]
\centering
\resizebox{\textwidth}{!}{%
\begin{tabular}{rlrlrlrl}
\multicolumn{2}{c}{TDNN-F HMM} & \multicolumn{2}{c}{ResNet + CTC + GCNN {\tiny(LIBRISPEECH)}} & \multicolumn{2}{c}{ResNet + CTC + 4-gram {\tiny(LIBRIVOX)}} & \multicolumn{2}{c}{TDS + CTC + 4-gram {\tiny(MyST)}} \\ 
\cmidrule(lr){1-2} \cmidrule(lr){3-4} \cmidrule(lr){5-6} \cmidrule(lr){7-8}
Frequency & Confusion & Frequency & Confusion & Frequency & Confusion & Frequency & Confusion\\
\cmidrule(lr){1-2} \cmidrule(lr){3-4} \cmidrule(lr){5-6} \cmidrule(lr){7-8}
933 & and $\rightarrow$ in              & 358 & and $\rightarrow$ an       & 434 & and $\rightarrow$ in         & 2281 & $<unk>$ $\rightarrow$ um\\
819 & and $\rightarrow$ an              & 311 & and $\rightarrow$ in       & 396 & $<unk>$ $\rightarrow$ m      & 230 & v $\rightarrow$ b\\
312 & uhm $\rightarrow$ arm             & 273 & r $\rightarrow$ are        & 290 & q $\rightarrow$ k            & 213 & n $\rightarrow$ and\\
197 & uhm $\rightarrow$ own             & 265 & q $\rightarrow$ you        & 285 & q $\rightarrow$ u            & 173 & r $\rightarrow$ are\\
186 & uhm $\rightarrow$ um              & 235 & $<unk>$ $\rightarrow$ and  & 251 & $<unk>$ $\rightarrow$ and    & 170 & a $\rightarrow$ the\\
169 & uhm $\rightarrow$ hum             & 211 & s $\rightarrow$ as         & 234 & $<unk>$ $\rightarrow$ a      & 153 & in $\rightarrow$ and\\
145 & uh $\rightarrow$ ah               & 186 & $<unk>$ $\rightarrow$ on   & 210 & a $\rightarrow$ the          & 145 & q $\rightarrow$ you\\
143 & the $\rightarrow$ a               & 183 & $<unk>$ $\rightarrow$ um   & 202 & v $\rightarrow$ b            & 145 & z $\rightarrow$ c\\
136 & uhm $\rightarrow$ oh              & 181 & v $\rightarrow$ the        & 192 & uh $\rightarrow$ a           & 139 & v $\rightarrow$ the\\
131 & uhm $\rightarrow$ am              & 177 & a $\rightarrow$ the        & 154 & gonna $\rightarrow$ to       & 138 & because $\rightarrow$ cause\\
129 & w $\rightarrow$ u                 & 168 & n $\rightarrow$ an         & 146 & $<unk>$ $\rightarrow$ um     & 136 & m $\rightarrow$ and\\
128 & uhm $\rightarrow$ on              & 164 & $<unk>$ $\rightarrow$ oh   & 137 & $<unk>$ $\rightarrow$ o      & 136 & u $\rightarrow$ you\\
124 & a $\rightarrow$ the               & 162 & uh $\rightarrow$ a         & 136 & $<unk>$ $\rightarrow$ on     & 130 & mom $\rightarrow$ ma\\
117 & uhm $\rightarrow$ m               & 144 & $<unk>$ $\rightarrow$ i    & 129 & $<unk>$ $\rightarrow$ ah     & 103 & and $\rightarrow$ in\\
104 & a $\rightarrow$ e                 & 139 & i $\rightarrow$ a          & 126 & $<unk>$ $\rightarrow$ oh     &  95 & $<unk>$ $\rightarrow$ and\\
100 & to $\rightarrow$ a                & 138 & $<unk>$ $\rightarrow$ a    & 119 & i $\rightarrow$ a            &  88 & to $\rightarrow$ the\\
100 & to $\rightarrow$ the              & 119 & m $\rightarrow$ an         & 115 & z $\rightarrow$ c            &  79 & gonna $\rightarrow$ to\\
94 & i $\rightarrow$ a                  & 99 & the $\rightarrow$ a         & 114 & v $\rightarrow$ e            &  77 & is $\rightarrow$ there's\\
91 & you $\rightarrow$ ye               & 99 & y $\rightarrow$ why         & 101 & and $\rightarrow$ n          &  76 & uh $\rightarrow$ um\\
89 & uh $\rightarrow$ er                & 92 & u $\rightarrow$ you         &  97 & the $\rightarrow$ a          &  75 & p $\rightarrow$ b\\
77 & n $\rightarrow$ an                 & 81 & she $\rightarrow$ he        &  92 & $<unk>$ $\rightarrow$ of     &  74 & z $\rightarrow$ the\\
75 & and $\rightarrow$ anne             & 77 & them $\rightarrow$ em       &  92 & mom $\rightarrow$ ma         &  73 & the $\rightarrow$ a\\
74 & it's $\rightarrow$ its             & 72 & gonna $\rightarrow$ to      &  86 & $<unk>$ $\rightarrow$ i      &  71 & he $\rightarrow$ you\\
73 & u $\rightarrow$ you                & 72 & z $\rightarrow$ why         &  84 & uh $\rightarrow$ ah          &  67 & v $\rightarrow$ you\\
73 & z $\rightarrow$ c                  & 71 & this $\rightarrow$ the      &  81 & yeah $\rightarrow$ yes       &  60 & this $\rightarrow$ the\\
72 & and $\rightarrow$ an'              & 70 & $<unk>$ $\rightarrow$ an    &  74 & in $\rightarrow$ and         &  58 & she $\rightarrow$ he\\
72 & and $\rightarrow$ on               & 69 & to $\rightarrow$ the        &  72 & i $\rightarrow$ j            &  56 & i $\rightarrow$ it\\
72 & uhm $\rightarrow$ of               & 69 & v $\rightarrow$ b           &  71 & to $\rightarrow$ the         &  54 & sister $\rightarrow$ system\\
70 & and $\rightarrow$ m                & 65 & to $\rightarrow$ a          &  69 & z $\rightarrow$ e            &  53 & is $\rightarrow$ he's\\
69 & a $\rightarrow$ eh                 & 64 & q $\rightarrow$ u           &  68 & $<unk>$ $\rightarrow$ i'm    &  53 & is $\rightarrow$ she's\\
68 & uhm $\rightarrow$ ah               & 62 & u $\rightarrow$ to          &  68 & yeah $\rightarrow$ ye        &  53 & my $\rightarrow$ like\\
67 & v $\rightarrow$ b                  & 60 & yeah $\rightarrow$ yes      &  65 & them $\rightarrow$ em        &  52 & uh $\rightarrow$ a\\
65 & a $\rightarrow$ of                 & 55 & $<unk>$ $\rightarrow$ i'm   &  64 & to $\rightarrow$ a           &  49 & $<unk>$ $\rightarrow$ the\\
65 & and $\rightarrow$ than             & 55 & in $\rightarrow$ and        &  61 & g $\rightarrow$ t            &  46 & dad $\rightarrow$ that\\
65 & uhm $\rightarrow$ an               & 52 & and $\rightarrow$ i         &  61 & is $\rightarrow$ there's     &  45 & g $\rightarrow$ d\\
63 & and $\rightarrow$ a                & 52 & is $\rightarrow$ there's    &  58 & this $\rightarrow$ the       &  44 & i $\rightarrow$ a\\
63 & and $\rightarrow$ eh               & 52 & v $\rightarrow$ you         &  57 & f $\rightarrow$ n            &  43 & i $\rightarrow$ um\\
62 & and $\rightarrow$ the              & 51 & and $\rightarrow$ why       &  57 & she $\rightarrow$ he         &  43 & uh $\rightarrow$ the\\
58 & and $\rightarrow$ then             & 51 & z $\rightarrow$ see         &  55 & $<unk>$ $\rightarrow$ the    &  42 & q $\rightarrow$ b\\
57 & like $\rightarrow$ liked           & 50 & and $\rightarrow$ the       &  54 & i $\rightarrow$ and          &  41 & he $\rightarrow$ it\\
56 & is $\rightarrow$ as                & 50 & is $\rightarrow$ he's       &  52 & and $\rightarrow$ m          &  41 & it's $\rightarrow$ is\\
56 & it's $\rightarrow$ is              & 50 & s $\rightarrow$ are         &  51 & $<unk>$ $\rightarrow$ in     &  39 & am $\rightarrow$ i'm\\
56 & t $\rightarrow$ to                 & 49 & $<unk>$ $\rightarrow$ am    &  51 & okay $\rightarrow$ k         &  39 & gonna $\rightarrow$ gon\\
55 & and $\rightarrow$ end              & 49 & $<unk>$ $\rightarrow$ em    &  51 & y $\rightarrow$ w            &  39 & u $\rightarrow$ d\\
54 & them $\rightarrow$ em              & 49 & r $\rightarrow$ you         &  49 & and $\rightarrow$ the        &  38 & and $\rightarrow$ um\\
54 & this $\rightarrow$ the             & 49 & t $\rightarrow$ are         &  49 & because $\rightarrow$ cause  &  37 & $<unk>$ $\rightarrow$ that\\
52 & favorite $\rightarrow$ favourite   & 49 & uh $\rightarrow$ ah         &  48 & and $\rightarrow$ a          &  37 & and $\rightarrow$ then\\
52 & two $\rightarrow$ to               & 48 & i $\rightarrow$ j           &  48 & z $\rightarrow$ w            &  37 & d $\rightarrow$ the\\
50 & yeah $\rightarrow$ yes             & 48 & q $\rightarrow$ e           &  47 & that's $\rightarrow$ that    &  37 & gonna $\rightarrow$ on\\
48 & i $\rightarrow$ j                  & 48 & s $\rightarrow$ you         &  47 & them $\rightarrow$ him       &  37 & lake $\rightarrow$ like\\
\bottomrule
\end{tabular}}
\caption{Top 50 Confusion Pairs: On OGI Corpus for top performing DNN-HMM system and end-to-end ASR systems}\label{tab:cp_ogi}
\end{table*}

\section{Conclusions}\label{sec:conclusion}
In this work, we presented a detailed empirical study of children speech recognition with the state-of-the-art end-to-end architectures.
The findings of the study suggest that the end-to-end speech recognition system trained on adult speech have short-comings for recognizing children speech.
In terms of WER, the children speech recognition with MyST corpus is approximately 10 times worse and on OGI Kids Corpus the performance is approximately 19 times worse compared to adult speech recognition.
With the end-to-end systems the gap in performance between adult and children is wider in comparison with the DNN-HMM hybrid systems, although in terms of absolute WER the end-to-end systems are a significant improvement over the latter.
The benefits established with end-to-end ASR for adult speech do not translate completely to children speech.
End-to-end architectures trained on large amounts of adult speech data can certainly help performance on children speech.
Addition of large amounts of adult speech is found to benefit more when the acoustic mismatch is large between children and adults.
Although, adaptation of acoustic model on children speech helps, the recognition performance remains more than 6 times worse compared to adult ASR.
DNN-HMM hybrid models benefit to a larger extent with children speech adaptation compared to end-to-end ASR, but the latter performs better in absolute WER.
Transformer network architectures are the best performing models when the train-test mismatch is low, however they do not generalize well when train-test mismatch is high including children age disparities.
CTC loss based models are robust to children speech recognition, however the sequence-to-sequence models can break down completely during high mismatch conditions with children speech recognition.
Our experiments indicate better performance with greedy decoding without language model for children ASR suggesting that acoustic mismatch dominates performance drop.

Insights into the errors reveal the end-to-end system have lower substitutions and insertions and high deletions on children speech recognition compared to hybrid DNN-HMM.
ASR of younger children still remains a challenge with end-to-end systems while the performance increases with increase in age similar to trends observed in GMM-HMM and DNN-HMM systems in prior literature.
On adaptation with children speech, the end-to-end systems provide near constant improvements over all age categories irrespective of age demographics of the adaptation data.
However, the DNN-HMM hybrid systems are more sensitive to age, giving skewed performance benefits for matched train-test age categories.
Training end-to-end systems with large amount of adult speech data benefits recognition for all age categories and younger children benefit to a greater degree.
End-to-end systems suffer in decoding longer utterances and specifically sequence-to-sequence models undergo drastic degradation compared to CTC models, whereas the DNN-HMM hybrid systems do not undergo any degradation.

Overall, the state-of-the-art end-to-end systems setting high benchmarks on adult speech are still far from achieving the same levels of performance for children speech.
This emphasizes the need to include children speech for developing benchmark tasks for ASR.
The results also point to fundamental challenges that still need to be addressed in children speech recognition  with end-to-end architectures.

\bibliographystyle{IEEEbib}
\bibliography{refs}

\begin{thebibliography}{10}

\bibitem{Narayanan2002Creatingconversationalinterfacesfor}
Shrikanth~S. Narayanan and Alexandros Potamianos,
\newblock ``Creating conversational interfaces for children,''
\newblock {\em IEEE Transactions on Speech and Audio Processing}, vol. 10, no.
  2, pp. 65--78, feb 2002.

\bibitem{Bone2017SignalProcessingandMachine}
Daniel Bone, Chi-Chun Lee, Theodora Chaspari, James Gibson, and Shrikanth
  Narayanan,
\newblock ``Signal processing and machine learning for mental health research
  and clinical applications,''
\newblock {\em IEEE Signal Processing Magazine}, vol. 34, no. 5, pp. 189--196,
  September 2017.

\bibitem{bone2017behavioral}
Daniel Bone, Theodora Chaspari, and Shrikanth Narayanan,
\newblock ``Behavioral signal processing and autism: Learning from multimodal
  behavioral signals,''
\newblock {\em Autism Imaging and Devices}, pp. 335--360, 2017.

\bibitem{dahl2011context}
George~E Dahl, Dong Yu, Li~Deng, and Alex Acero,
\newblock ``Context-dependent pre-trained deep neural networks for
  large-vocabulary speech recognition,''
\newblock {\em IEEE Transactions on audio, speech, and language processing},
  vol. 20, no. 1, pp. 30--42, 2011.

\bibitem{graves2006connectionist}
Alex Graves, Santiago Fern{\'a}ndez, Faustino Gomez, and J{\"u}rgen
  Schmidhuber,
\newblock ``Connectionist temporal classification: labelling unsegmented
  sequence data with recurrent neural networks,''
\newblock in {\em Proceedings of the 23rd international conference on Machine
  learning}, 2006, pp. 369--376.

\bibitem{sutskever2014sequence}
Ilya Sutskever, Oriol Vinyals, and Quoc~V Le,
\newblock ``Sequence to sequence learning with neural networks,''
\newblock in {\em Advances in neural information processing systems}, 2014, pp.
  3104--3112.

\bibitem{chorowski2015attention}
Jan~K Chorowski, Dzmitry Bahdanau, Dmitriy Serdyuk, Kyunghyun Cho, and Yoshua
  Bengio,
\newblock ``Attention-based models for speech recognition,''
\newblock in {\em Advances in neural information processing systems}, 2015, pp.
  577--585.

\bibitem{chan2016listen}
William Chan, Navdeep Jaitly, Quoc Le, and Oriol Vinyals,
\newblock ``Listen, attend and spell: A neural network for large vocabulary
  conversational speech recognition,''
\newblock in {\em 2016 IEEE International Conference on Acoustics, Speech and
  Signal Processing (ICASSP)}. IEEE, 2016, pp. 4960--4964.

\bibitem{chiu2018state}
Chung-Cheng Chiu, Tara~N Sainath, Yonghui Wu, Rohit Prabhavalkar, Patrick
  Nguyen, Zhifeng Chen, Anjuli Kannan, Ron~J Weiss, Kanishka Rao, Ekaterina
  Gonina, et~al.,
\newblock ``State-of-the-art speech recognition with sequence-to-sequence
  models,''
\newblock in {\em 2018 IEEE International Conference on Acoustics, Speech and
  Signal Processing (ICASSP)}. IEEE, 2018, pp. 4774--4778.

\bibitem{zhang2017towards}
Ying Zhang, Mohammad Pezeshki, Phil{\'e}mon Brakel, Saizheng Zhang, Cesar
  Laurent~Yoshua Bengio, and Aaron Courville,
\newblock ``Towards end-to-end speech recognition with deep convolutional
  neural networks,''
\newblock {\em arXiv preprint arXiv:1701.02720}, 2017.

\bibitem{zeghidour2018fully}
Neil Zeghidour, Qiantong Xu, Vitaliy Liptchinsky, Nicolas Usunier, Gabriel
  Synnaeve, and Ronan Collobert,
\newblock ``Fully convolutional speech recognition,''
\newblock {\em arXiv preprint arXiv:1812.06864}, 2018.

\bibitem{pmlr-v48-amodei16}
Dario Amodei, Sundaram Ananthanarayanan, Rishita Anubhai, Jingliang Bai, Eric
  Battenberg, Carl Case, Jared Casper, Bryan Catanzaro, Qiang Cheng, Guoliang
  Chen, Jie Chen, Jingdong Chen, Zhijie Chen, Mike Chrzanowski, Adam Coates,
  Greg Diamos, Ke~Ding, Niandong Du, Erich Elsen, Jesse Engel, Weiwei Fang,
  Linxi Fan, Christopher Fougner, Liang Gao, Caixia Gong, Awni Hannun, Tony
  Han, Lappi Johannes, Bing Jiang, Cai Ju, Billy Jun, Patrick LeGresley, Libby
  Lin, Junjie Liu, Yang Liu, Weigao Li, Xiangang Li, Dongpeng Ma, Sharan
  Narang, Andrew Ng, Sherjil Ozair, Yiping Peng, Ryan Prenger, Sheng Qian,
  Zongfeng Quan, Jonathan Raiman, Vinay Rao, Sanjeev Satheesh, David Seetapun,
  Shubho Sengupta, Kavya Srinet, Anuroop Sriram, Haiyuan Tang, Liliang Tang,
  Chong Wang, Jidong Wang, Kaifu Wang, Yi~Wang, Zhijian Wang, Zhiqian Wang,
  Shuang Wu, Likai Wei, Bo~Xiao, Wen Xie, Yan Xie, Dani Yogatama, Bin Yuan, Jun
  Zhan, and Zhenyao Zhu,
\newblock ``Deep speech 2 : End-to-end speech recognition in english and
  mandarin,''
\newblock in {\em Proceedings of The 33rd International Conference on Machine
  Learning}, Maria~Florina Balcan and Kilian~Q. Weinberger, Eds., New York, New
  York, USA, 20--22 Jun 2016, vol.~48 of {\em Proceedings of Machine Learning
  Research}, pp. 173--182, PMLR.

\bibitem{zhang2017very}
Yu~Zhang, William Chan, and Navdeep Jaitly,
\newblock ``Very deep convolutional networks for end-to-end speech
  recognition,''
\newblock in {\em 2017 IEEE International Conference on Acoustics, Speech and
  Signal Processing (ICASSP)}. IEEE, 2017, pp. 4845--4849.

\bibitem{wang2017residual}
Yisen Wang, Xuejiao Deng, Songbai Pu, and Zhiheng Huang,
\newblock ``Residual convolutional ctc networks for automatic speech
  recognition,''
\newblock {\em arXiv preprint arXiv:1702.07793}, 2017.

\bibitem{kim2017residual}
Jaeyoung Kim, Mostafa El-Khamy, and Jungwon Lee,
\newblock ``Residual lstm: Design of a deep recurrent architecture for distant
  speech recognition,''
\newblock {\em arXiv preprint arXiv:1701.03360}, 2017.

\bibitem{pundak2017highway}
Golan Pundak and Tara~N Sainath,
\newblock ``Highway-lstm and recurrent highway networks for speech
  recognition,''
\newblock {\em Proc. Interspeech 2017}, pp. 1303--1307, 2017.

\bibitem{watanabe2017hybrid}
Shinji Watanabe, Takaaki Hori, Suyoun Kim, John~R Hershey, and Tomoki Hayashi,
\newblock ``Hybrid ctc/attention architecture for end-to-end speech
  recognition,''
\newblock {\em IEEE Journal of Selected Topics in Signal Processing}, vol. 11,
  no. 8, pp. 1240--1253, 2017.

\bibitem{kim2017joint}
Suyoun Kim, Takaaki Hori, and Shinji Watanabe,
\newblock ``Joint ctc-attention based end-to-end speech recognition using
  multi-task learning,''
\newblock in {\em 2017 IEEE international conference on acoustics, speech and
  signal processing (ICASSP)}. IEEE, 2017, pp. 4835--4839.

\bibitem{dong2018speech}
Linhao Dong, Shuang Xu, and Bo~Xu,
\newblock ``Speech-transformer: a no-recurrence sequence-to-sequence model for
  speech recognition,''
\newblock in {\em 2018 IEEE International Conference on Acoustics, Speech and
  Signal Processing (ICASSP)}. IEEE, 2018, pp. 5884--5888.

\bibitem{synnaeve2019end}
Gabriel Synnaeve, Qiantong Xu, Jacob Kahn, Edouard Grave, Tatiana Likhomanenko,
  Vineel Pratap, Anuroop Sriram, Vitaliy Liptchinsky, and Ronan Collobert,
\newblock ``End-to-end asr: from supervised to semi-supervised learning with
  modern architectures,''
\newblock {\em arXiv preprint arXiv:1911.08460}, 2019.

\bibitem{lee1999acoustics}
Sungbok Lee, Alexandros Potamianos, and Shrikanth Narayanan,
\newblock ``Acoustics of children’s speech: Developmental changes of temporal
  and spectral parameters,''
\newblock {\em The Journal of the Acoustical Society of America}, vol. 105, no.
  3, pp. 1455--1468, 1999.

\bibitem{Lee2014Developmentalacousticstudyof}
Sungbok Lee, Alexandros Potamianos, and Shrikanth~S. Narayanan,
\newblock ``Developmental acoustic study of american english diphthongs,''
\newblock {\em J. Acoust. Soc. Am.}, vol. 136, no. 4, pp. 1880--1894, oct 2014.

\bibitem{potamianos2003robust}
Alexandros Potamianos and Shrikanth Narayanan,
\newblock ``Robust recognition of children's speech,''
\newblock {\em IEEE Transactions on speech and audio processing}, vol. 11, no.
  6, pp. 603--616, 2003.

\bibitem{gerosa2007acoustic}
Matteo Gerosa, Diego Giuliani, and Fabio Brugnara,
\newblock ``Acoustic variability and automatic recognition of children’s
  speech,''
\newblock {\em Speech Communication}, vol. 49, no. 10-11, pp. 847--860, 2007.

\bibitem{potamianos1997automatic}
Alexandros Potamianos, Shrikanth Narayanan, and Sungbok Lee,
\newblock ``Automatic speech recognition for children,''
\newblock in {\em Fifth European Conference on Speech Communication and
  Technology}, 1997, pp. 2371--2374.

\bibitem{gallagher1977revision}
Tanya~M Gallagher,
\newblock ``Revision behaviors in the speech of normal children developing
  language,''
\newblock {\em Journal of Speech and Hearing Research}, vol. 20, no. 2, pp.
  303--318, 1977.

\bibitem{ghai2009exploring}
Shweta Ghai and Rohit Sinha,
\newblock ``Exploring the role of spectral smoothing in context of children's
  speech recognition,''
\newblock in {\em Tenth Annual Conference of the International Speech
  Communication Association}, 2009, pp. 1607--1610.

\bibitem{sinha2018assessment}
Rohit Sinha and Syed Shahnawazuddin,
\newblock ``Assessment of pitch-adaptive front-end signal processing for
  children’s speech recognition,''
\newblock {\em Computer Speech \& Language}, vol. 48, pp. 103--121, 2018.

\bibitem{shivakumar2014improving}
Prashanth~Gurunath Shivakumar, Alexandros Potamianos, Sungbok Lee, and
  Shrikanth~S Narayanan,
\newblock ``Improving speech recognition for children using acoustic adaptation
  and pronunciation modeling.,''
\newblock in {\em WOCCI}, 2014, pp. 15--19.

\bibitem{giuliani2003investigating}
Diego Giuliani and Matteo Gerosa,
\newblock ``Investigating recognition of children's speech,''
\newblock in {\em 2003 IEEE International Conference on Acoustics, Speech, and
  Signal Processing, 2003. Proceedings.(ICASSP'03).} IEEE, 2003, vol.~2, pp.
  II--137.

\bibitem{giuliani2006improved}
Diego Giuliani, Matteo Gerosa, and Fabio Brugnara,
\newblock ``Improved automatic speech recognition through speaker
  normalization,''
\newblock {\em Computer Speech \& Language}, vol. 20, no. 1, pp. 107--123,
  2006.

\bibitem{shivakumar2020transfer}
Prashanth~Gurunath Shivakumar and Panayiotis Georgiou,
\newblock ``Transfer learning from adult to children for speech recognition:
  Evaluation, analysis and recommendations,''
\newblock {\em Computer Speech \& Language}, vol. 63, pp. 101077, 2020.

\bibitem{gales1998maximum}
Mark~JF Gales,
\newblock ``Maximum likelihood linear transformations for hmm-based speech
  recognition,''
\newblock {\em Computer speech \& language}, vol. 12, no. 2, pp. 75--98, 1998.

\bibitem{li2002analysis}
Qun Li and Martin~J Russell,
\newblock ``An analysis of the causes of increased error rates in children's
  speech recognition,''
\newblock in {\em Seventh International Conference on Spoken Language
  Processing}, 2002, pp. 2337--2340.

\bibitem{das1998improvements}
Subrata Das, Don Nix, and Michael Picheny,
\newblock ``Improvements in children's speech recognition performance,''
\newblock in {\em Proceedings of the 1998 IEEE International Conference on
  Acoustics, Speech and Signal Processing, ICASSP'98 (Cat. No. 98CH36181)}.
  IEEE, 1998, vol.~1, pp. 433--436.

\bibitem{ng2020cuhk}
Si-Ioi Ng, Wei Liu, Zhiyuan Peng, Siyuan Feng, Hing-Pang Huang, Odette
  Scharenborg, and Tan Lee,
\newblock ``The cuhk-tudelft system for the slt 2021 children speech
  recognition challenge,''
\newblock {\em arXiv preprint arXiv:2011.06239}, 2020.

\bibitem{chen2020data}
Guoguo Chen, Xingyu Na, Yongqing Wang, Zhiyong Yan, Junbo Zhang, Sifan Ma, and
  Yujun Wang,
\newblock ``Data augmentation for children's speech recognition--the"
  ethiopian" system for the slt 2021 children speech recognition challenge,''
\newblock {\em arXiv preprint arXiv:2011.04547}, 2020.

\bibitem{yu2020slt}
Fan Yu, Zhuoyuan Yao, Xiong Wang, Keyu An, Lei Xie, Zhijian Ou, Bo~Liu, Xiulin
  Li, and Guanqiong Miao,
\newblock ``The slt 2021 children speech recognition challenge: Open datasets,
  rules and baselines,''
\newblock {\em arXiv preprint arXiv:2011.06724}, 2020.

\bibitem{povey2018semi}
Daniel Povey, Gaofeng Cheng, Yiming Wang, Ke~Li, Hainan Xu, Mahsa Yarmohammadi,
  and Sanjeev Khudanpur,
\newblock ``Semi-orthogonal low-rank matrix factorization for deep neural
  networks.,''
\newblock in {\em Interspeech}, 2018, pp. 3743--3747.

\bibitem{wu2019advances}
Fei Wu, Leibny~Paola Garc{\'\i}a-Perera, Daniel Povey, and Sanjeev Khudanpur,
\newblock ``Advances in automatic speech recognition for child speech using
  factored time delay neural network.,''
\newblock in {\em INTERSPEECH}, 2019, pp. 1--5.

\bibitem{povey2016purely}
Daniel Povey, Vijayaditya Peddinti, Daniel Galvez, Pegah Ghahremani, Vimal
  Manohar, Xingyu Na, Yiming Wang, and Sanjeev Khudanpur,
\newblock ``Purely sequence-trained neural networks for asr based on
  lattice-free mmi.,''
\newblock in {\em Interspeech}, 2016, pp. 2751--2755.

\bibitem{he2015deep}
Kaiming He, Xiangyu Zhang, Shaoqing Ren, and Jian Sun,
\newblock ``Deep residual learning for image recognition,'' 2015.

\bibitem{xiong2016achieving}
Wayne Xiong, Jasha Droppo, Xuedong Huang, Frank Seide, Mike Seltzer, Andreas
  Stolcke, Dong Yu, and Geoffrey Zweig,
\newblock ``Achieving human parity in conversational speech recognition,''
\newblock {\em arXiv preprint arXiv:1610.05256}, 2016.

\bibitem{saon2017english}
George Saon, Gakuto Kurata, Tom Sercu, Kartik Audhkhasi, Samuel Thomas,
  Dimitrios Dimitriadis, Xiaodong Cui, Bhuvana Ramabhadran, Michael Picheny,
  Lynn-Li Lim, et~al.,
\newblock ``English conversational telephone speech recognition by humans and
  machines,''
\newblock {\em arXiv preprint arXiv:1703.02136}, 2017.

\bibitem{park2019specaugment}
Daniel~S Park, William Chan, Yu~Zhang, Chung-Cheng Chiu, Barret Zoph, Ekin~D
  Cubuk, and Quoc~V Le,
\newblock ``Specaugment: A simple data augmentation method for automatic speech
  recognition,''
\newblock {\em arXiv preprint arXiv:1904.08779}, 2019.

\bibitem{ba2016layer}
Jimmy~Lei Ba, Jamie~Ryan Kiros, and Geoffrey~E Hinton,
\newblock ``Layer normalization,''
\newblock {\em arXiv preprint arXiv:1607.06450}, 2016.

\bibitem{hannun2019sequence}
Awni Hannun, Ann Lee, Qiantong Xu, and Ronan Collobert,
\newblock ``Sequence-to-sequence speech recognition with time-depth separable
  convolutions,''
\newblock {\em arXiv preprint arXiv:1904.02619}, 2019.

\bibitem{vaswani2017attention}
Ashish Vaswani, Noam Shazeer, Niki Parmar, Jakob Uszkoreit, Llion Jones,
  Aidan~N Gomez, {\L}ukasz Kaiser, and Illia Polosukhin,
\newblock ``Attention is all you need,''
\newblock in {\em Advances in neural information processing systems}, 2017, pp.
  5998--6008.

\bibitem{devlin2018bert}
Jacob Devlin, Ming-Wei Chang, Kenton Lee, and Kristina Toutanova,
\newblock ``Bert: Pre-training of deep bidirectional transformers for language
  understanding,''
\newblock {\em arXiv preprint arXiv:1810.04805}, 2018.

\bibitem{karita2019comparative}
Shigeki Karita, Nanxin Chen, Tomoki Hayashi, Takaaki Hori, Hirofumi Inaguma,
  Ziyan Jiang, Masao Someki, Nelson Enrique~Yalta Soplin, Ryuichi Yamamoto,
  Xiaofei Wang, et~al.,
\newblock ``A comparative study on transformer vs rnn in speech applications,''
\newblock in {\em 2019 IEEE Automatic Speech Recognition and Understanding
  Workshop (ASRU)}. IEEE, 2019, pp. 449--456.

\bibitem{parmar2018image}
Niki Parmar, Ashish Vaswani, Jakob Uszkoreit, {\L}ukasz Kaiser, Noam Shazeer,
  Alexander Ku, and Dustin Tran,
\newblock ``Image transformer,''
\newblock {\em arXiv preprint arXiv:1802.05751}, 2018.

\bibitem{fan2019reducing}
Angela Fan, Edouard Grave, and Armand Joulin,
\newblock ``Reducing transformer depth on demand with structured dropout,''
\newblock {\em arXiv preprint arXiv:1909.11556}, 2019.

\bibitem{dauphin2017language}
Yann~N Dauphin, Angela Fan, Michael Auli, and David Grangier,
\newblock ``Language modeling with gated convolutional networks,''
\newblock in {\em International conference on machine learning}. PMLR, 2017,
  pp. 933--941.

\bibitem{likhomanenko2019needs}
Tatiana Likhomanenko, Gabriel Synnaeve, and Ronan Collobert,
\newblock ``Who needs words? lexicon-free speech recognition,''
\newblock {\em arXiv preprint arXiv:1904.04479}, 2019.

\bibitem{ward2011my}
Wayne Ward, Ronald Cole, Daniel Bola{\~n}os, Cindy Buchenroth-Martin, Edward
  Svirsky, Sarel~Van Vuuren, Timothy Weston, Jing Zheng, and Lee Becker,
\newblock ``My science tutor: A conversational multimedia virtual tutor for
  elementary school science,''
\newblock {\em ACM Transactions on Speech and Language Processing (TSLP)}, vol.
  7, no. 4, pp. 1--29, 2011.

\bibitem{ward2019my}
Wayne Ward, Ron Cole, and Sameer Pradhan,
\newblock ``My science tutor and the myst corpus,''
\newblock 2019.

\bibitem{shobaki2000ogi}
Khaldoun Shobaki, John-Paul Hosom, and Ronald~A Cole,
\newblock ``The ogi kids' speech corpus and recognizers,''
\newblock in {\em Sixth International Conference on Spoken Language
  Processing}, 2000, vol.~4, pp. 258--261.

\bibitem{gerosa2006acoustic}
Matteo Gerosa, Diego Giuliani, and Shrikanth Narayanan,
\newblock ``Acoustic analysis and automatic recognition of spontaneous
  children's speech,''
\newblock in {\em Ninth International Conference on Spoken Language
  Processing}, 2006.

\bibitem{povey2011kaldi}
Daniel Povey, Arnab Ghoshal, Gilles Boulianne, Lukas Burget, Ondrej Glembek,
  Nagendra Goel, Mirko Hannemann, Petr Motlicek, Yanmin Qian, Petr Schwarz,
  et~al.,
\newblock ``The kaldi speech recognition toolkit,''
\newblock in {\em IEEE 2011 workshop on automatic speech recognition and
  understanding}. IEEE Signal Processing Society, 2011, number CONF.

\bibitem{panayotov2015librispeech}
Vassil Panayotov, Guoguo Chen, Daniel Povey, and Sanjeev Khudanpur,
\newblock ``Librispeech: an asr corpus based on public domain audio books,''
\newblock in {\em 2015 IEEE International Conference on Acoustics, Speech and
  Signal Processing (ICASSP)}. IEEE, 2015, pp. 5206--5210.

\bibitem{pratap2018wav2letter++}
Vineel Pratap, Awni Hannun, Qiantong Xu, Jeff Cai, Jacob Kahn, Gabriel
  Synnaeve, Vitaliy Liptchinsky, and Ronan Collobert,
\newblock ``wav2letter++: The fastest open-source speech recognition system,''
\newblock {\em arXiv preprint arXiv:1812.07625}, 2018.

\bibitem{kudo2018sentencepiece}
Taku Kudo and John Richardson,
\newblock ``Sentencepiece: A simple and language independent subword tokenizer
  and detokenizer for neural text processing,''
\newblock {\em arXiv preprint arXiv:1808.06226}, 2018.

\bibitem{heafield2011kenlm}
Kenneth Heafield,
\newblock ``Kenlm: Faster and smaller language model queries,''
\newblock in {\em Proceedings of the sixth workshop on statistical machine
  translation}, 2011, pp. 187--197.

\bibitem{ott2019fairseq}
Myle Ott, Sergey Edunov, Alexei Baevski, Angela Fan, Sam Gross, Nathan Ng,
  David Grangier, and Michael Auli,
\newblock ``fairseq: A fast, extensible toolkit for sequence modeling,''
\newblock {\em arXiv preprint arXiv:1904.01038}, 2019.

\bibitem{zeghidour2019fully}
Neil Zeghidour, Qiantong Xu, Vitaliy Liptchinsky, Nicolas Usunier, Gabriel
  Synnaeve, and Ronan Collobert,
\newblock ``Fully convolutional speech recognition,'' 2019.

\end{thebibliography}

\end{document}